\documentclass[11pt, oneside, a4paper]{article}

\usepackage[left=2cm,right=2cm,top=2cm,bottom=2.4cm]{geometry}

\usepackage{amsthm,amsmath,natbib}
\RequirePackage[colorlinks,citecolor=blue,urlcolor=blue]{hyperref}

\usepackage{xcolor}
\usepackage{csquotes}

\usepackage{amsfonts}
\usepackage{bm}
\usepackage{enumitem}

\usepackage{authblk}   
\usepackage{setspace}
\usepackage{totcount}
\newtotcounter{citnum} 
\def\oldbibitem{} \let\oldbibitem=\bibitem
\def\bibitem{\stepcounter{citnum}\oldbibitem}

\usepackage{ragged2e}
\usepackage{tikz} 
\usetikzlibrary{chains,shapes.arrows,fit}

\pgfdeclarelayer{background}
\pgfsetlayers{background,main}

\newcounter{task}

\newlength\taskwidth
\newlength\taskvsep

\def\taskpos{}
\def\taskanchor{}

\newcommand\task[1]{%
  {\parbox[t]{\taskwidth}{\scriptsize\Centering#1}}}

\tikzset{
inner/.style={
  on chain,
  circle,
  inner sep=2pt,
  fill=circlecolor,
  line width=1.5pt,
  draw=bordercolor,
  text width=2.5em,
  align=center,
  text height=1.25ex,
  text depth=0ex
},
on grid
}

\newcommand\Task[2][]{%
\node[inner xsep=0pt] (c1) {\phantom{1}};
\stepcounter{task}
\ifodd\thetask\relax
  \renewcommand\taskpos{\taskvsep}\renewcommand\taskanchor{south}
\else
  \renewcommand\taskpos{-\taskvsep}\renewcommand\taskanchor{north}
\fi
\node[inner,font=\footnotesize\sffamily\color{textcolor}]    
  (c\the\numexpr\value{task}+1\relax) {#1};
\node[anchor=\taskanchor,yshift=\taskpos] 
  at (c\the\numexpr\value{task}+1\relax) {\task{#2}};
}

\newcommand\drawarrow{
\ifnum\thetask=0\relax
  \node[on chain] (c1) {}; 
\fi
\node[on chain] (f) {};
\begin{pgfonlayer}{background}
\node[
  inner sep=10pt,
  single arrow,
  single arrow head extend=0.6cm,
  draw=none,
  fill=arrowcolor,
  fit= (c1) (f)
] (arrow) {};
\fill[white] 
  (arrow.before tail) -- (c1|-arrow.west) -- (arrow.after tail) -- cycle;
\end{pgfonlayer}
}

\newenvironment{timeline}[1][node distance=.75\taskwidth]
  {\par\noindent\begin{tikzpicture}[start chain,#1]}
  {\drawarrow\end{tikzpicture}\par}
  
\newcommand*{\changes}[1]{#1}
\newcommand*{\secondround}[1]{#1}
\newcommand*{\thirdround}[1]{#1}

\title{\textbf{Response-adaptive randomization in clinical trials: from myths to practical considerations}}

\author[1]{David S.\ Robertson}
\author[1]{Kim May Lee}
\author[1]{Boryana C.\ L\'{o}pez-Kolkovska}
\author[1]{Sof\'{i}a S.~Villar \footnote{Address correspondence to Sof\'{i}a S.~Villar,  MRC Biostatistics Unit, University of Cambridge, East Forvie Site, Robinson Way, Cambridge CB2 0SR, UK; E-mail: sofia.villar@mrc-bsu.cam.ac.uk}}
\affil[1]{\small MRC Biostatistics Unit, University of Cambridge, Cambridge, UK}

\date{\vspace{-24pt}}

\begin{document}

\maketitle

\doublespacing

\begin{abstract}
Response-\thirdround{A}daptive \thirdround{R}andomization (RAR) is part of a wider class of data-dependent sampling algorithms, for which clinical trials \changes{are typically} used as a motivating application. In that context, patient allocation to treatments is determined by randomization probabilities that \secondround{change} based on the accrued response data in order to achieve experimental goals. RAR has received abundant theoretical attention from the biostatistical literature since the 1930's and has been the subject of \changes{numerous} debates. \changes{In the last decade,} it has received renewed consideration from the applied \changes{and methodological} communities, \changes{driven by} \secondround{well-known} practical examples and its widespread use in machine learning. Papers on the subject \changes{present different views on its usefulness, and \secondround{these are not easy to} reconcile.} This work aims to address \changes{this gap by providing a unified, broad} and fresh review of methodological and practical issues to consider when debating the use of RAR in clinical trials.\\[-6pt]

\end{abstract}


\noindent {\small \textbf{Keywords:} Ethics; patient allocation; power; sample size imbalance, time trends, type~I error control.}

\vspace{0.5cm}


\section{Introduction}
\label{sec:intro}

Randomization to allocate patients to treatments is 
a defining element of a well-conducted study, ensuring comparability of treatment groups, mitigating selection bias, and providing the basis for statistical inference~\citep{Rosenberger2016}. In clinical trials, a 
randomization \changes{scheme which \secondround{remains unchanged} with patient responses} is still the most \changes{frequently used patient allocation procedure}.
\changes{Alternatively, randomization probabilities can be \emph{adapted}} during the trial based on the accrued 
responses, with the aim of achieving experimental objectives. 
\secondround{Objectives that can be targeted with a \thirdround{Response-Adaptive Randomization} (RAR) procedure include maximizing}
power of a specific treatment comparison and assigning more patients to an effective treatment during the trial. 

Few topics in the biostatistical literature have \changes{received as much attention \secondround{over the years} as RAR (also known as outcome-adaptive randomization)}.
RAR has been a fertile area of methodological research,
\secondround{as illustrated by} the reference section of this paper.
Despite this, the uptake of RAR in \changes{clinical} trial practice remains disproportionately low in comparison with the theoretical \secondround{interest it has generated since 
first proposed by \cite{Thompson1933}.  Its value in clinical trials \changes{remains a subject of active debate} within biostatistics, 
especially} 
during health care crises such as the Ebola outbreak~\citep{Brittain2016, Berry2016} or the COVID-19 pandemic~\citep{Proschan2020, Magaret2020, Villar2020}. 

\changes{This continued conversation has been enriching, but \thirdround{is} also often presented in papers} geared towards \changes{arguing either in favor or against its use} in clinical trials, \secondround{which has given RAR a controversial flavour. As well, some of these debates have been somewhat repetitive, as seen by how many of the points raised by \citet{Armitage1985} over 35 years ago continue to be revisited}. \changes{Examples of possibly} conflicting views \secondround{on the use of RAR} are given below.


\begin{displayquote}
If you are planning a randomized comparative clinical trial and someone proposes that you use outcome adaptive randomization, Just Say No. \citep{Thall2020}
\end{displayquote}

\begin{displayquote}
\ldots\ optimal [RAR] designs allow implementation of complex optimal allocations in multiple-objective clinical trials and provide valid tools to inference in the end of the trial. In many instances they prove superior over traditional balanced randomization designs in terms of both statistical efficiency and ethical criteria. \citep{Rosenberger2012}
\end{displayquote}


\begin{displayquote}
\thirdround{RAR} is a noble attempt to increase the likelihood that patients receive better performing treatments, but it causes numerous problems that more than offset any potential benefits. We discourage the use of RAR in clinical trials. 
\citep{Proschan2020}
\end{displayquote}

\noindent \secondround{The above examples 
help explain why the use of RAR in clinical trials remains rare and debated. It also suggests that, given the many different classes of RAR that exist, making general statements around \thirdround{the} relative merits of RAR may well be an elusive goal. This paper therefore aims \thirdround{to give} a balanced and fresh perspective. Instead of conveying a position in favor or against the use of RAR in clinical trials in general, we emphasize the less commonly known arguments (which also tend to be ones that are more positive towards the use of RAR).}

In parallel \changes{and in stark contrast} to this \changes{discussion}, in machine learning the uptake and popularity of Bayesian RAR \changes{(BRAR)}, \secondround{also referred to as Thompson \thirdround{S}ampling (TS)}, has been incredibly high~\citep{Kaufmann2017, Kaibel2019, Lattimore2019}. Their use in practice has been driven by substantial gains in system performances. \changes{Meanwhile,} in the clinical trial community, a crucial 
development was the \changes{use of BRAR in} some well-known biomarker led trials such as I-SPY 2 \citep{Barker2009} or BATTLE \citep{Kim2011}. The goal of these trials was to learn which subgroups (if any) benefit from a therapy and to change the randomization to favor patient allocation in that direction. \secondround{While these trials include other elements besides RAR}, they have set a precedent that RAR is feasible (at least in oncology), and have set expectations which, contrary to what the ECMO trials did in the 1980s (see Section~\ref{sec:history}), are driving investigators towards RAR in \secondround{other} contexts\thirdround{.} 
Both in the machine learning literature and in these trials, 
the \changes{BRAR} methodology used 
is a \changes{sub}class of the larger family of \changes{RAR} methods. 

\changes{After an extensive review of the literature, we recognized the need for \secondround{an updated and} broad discussion} 
\secondround{aimed at reconciling apparently conflicting arguments. We believe this is important because} some of these (mostly negative) positions on RAR persist, despite recent methodological developments \changes{over the past 10 years} directly addressing past criticisms \secondround{(see for example Section~\ref{subsec:time_trends})}. \changes{We compare recently proposed RAR procedures and use a new simulation study (in Section~\ref{subsec:imbalance}) to illustrate \secondround{how some viewpoints can tell only part of the story while a broad look can change conclusions.} Additionally, we hope this paper will   
drive methodological research towards areas that are less developed and help those considering the use of RAR in a specific experiment to navigate the relevant literature in light of \changes{recent opposing} views~\citep{Proschan2020, Villar2020, Magaret2020}.} 
\secondround{Overall, our ultimate message is a call for careful thinking about how to best deliver experimental goals through the appropriate use of trial adaptations \changes{including (but not limited to)} RAR.  }

We \changes{end this section} by providing \changes{some general notation, basic concepts and metrics to assess RAR. We give a} historical overview of RAR \thirdround{in} Section~\ref{sec:history}\thirdround{,} \changes{including a summary of classification criteria of different procedures (Section~\ref{subsec:taxonomies})}. We \changes{subsequently} \thirdround{explore} \changes{some key established views} about RAR \thirdround{in} Section~\ref{sec:myths}. \thirdround{We conclude with} \secondround{final considerations and \thirdround{a} discussion} in Section~\ref{sec:discuss}. \\[-6pt]

\subsection{\changes{Some notation and basic concepts}}
\label{subsec:notation}

\changes{We first describe the setting and notation necessary for a rigorous presentation of the debate around RAR. \secondround{Note that Table~\ref{tab:acronyms} \thirdround{in the Appendix} provides a summary of all the acronyms used in this paper}. Our focus is on clinical trials in which \secondround{a fixed} number of experimental treatments (labeled $1, \ldots, K$ with $K \geq 1$) are compared against a control or standard of care treatment (labeled $0$) in a sample of~$n$ patients. 
\secondround{The sample size~$n$ is also assumed fixed. This can, in principle, be relaxed \thirdround{to allow for} early stopping \thirdround{of the trial,} but} for the purposes of this paper we consider early stopping as a distinct type of adaptation. When treatment $k \in \{0, 1, \ldots, K\}$ is assigned to patient~$i$ (for $i \in \{1, \dots,n\}$), this generates a random response variable $Y_{k,i}$, which represents the primary outcome measure of the clinical trial. }

\changes{We let $a_{k,i}$ be a binary indicator variable denoting the observed treatment allocation for patient~$i$, with $a_{k,i} = 1$ if patient~$i$ is allocated to treatment~$k$ and \thirdround{$a_{k,i}=0$} otherwise. Each patient is allocated to one treatment only, and hence $\sum_{k=0}^{K} a_{k,i} = 1$. Typically patients enter the trial and are treated sequentially, either individually or in groups. In most of the RAR literature, patients are assumed to be randomized and treated one after another, with each patient's outcome \thirdround{being} available before the next patient needs to be treated. This assumption can be relaxed 
and incorporate delayed patient outcomes (e.g.\ for time-to-event data).}

\changes{
We assume $Y_{k,i}$ depends on a treatment-specific parameter of interest $\theta_k$. 
For notational convenience, we let~$Y_i$ denote the \secondround{realised} outcome of patient~$i$. 
We assume a parametric model for the primary outcome, ignoring nuisance parameters and other parameters of secondary interest for the final analysis.
For example, one could have a Bernoulli model for binary responses, where $\theta_k = p_k$ (the probability of a successful outcome for a patient on treatment~$k$):
\begin{equation}
Pr(Y_{k,i}=y \, | \, a_{k,i}=1) = p_k^{y} (1-p_k)^{(1-y)} \qquad \text{for } y=0,1.
\end{equation}
Other examples include a normal or exponential model for continuous outcome variables.}

\changes{As a general way to represent treatment allocation rules, we let $\pi_{k,i} = P(a_{k,i} = 1)$ denote the probability that patient~$i$ is allocated treatment~$k$. Note that we require $\sum_{k=0}^{K} \pi_{k,i} = 1$ and $\pi_{k,i} > 0 \: \forall i$. 
\thirdround{Also} \secondround{note that} our definition excludes non-randomized response-adaptive methods like the Gittins Index \citep{Villar2015a}. Traditional (fixed) randomization has $\pi_{k,i} = c_k$ for all~$i$, and \secondround{for} implementing \thirdround{E}qual \thirdround{R}andomization (ER) we set $c_k = 1/(K+1)$ for all~$k$. Finally, we let~$N_k$ denote the total number of patients that are allocated to treatment~$k$ by the end of the trial. In general, $N_k=\sum_{i = 0}^n a_{k,i}$ is a random variable, with the constraint $\sum_{k = 0}^K N_k = n$.}

\changes{In a RAR procedure, the allocation probabilities that define the \emph{sampling strategy} are adapted based on the past treatment allocations and response data. More formally, let $\bm{a_i} = (a_{0,i} , \, a_{1,i} , \ldots , \, a_{K,i})$ denote the allocation vector for patient~$i$. We also let $\bm{a}^{(j)} = \{\bm{a_1} , \ldots , \, \bm{a_j}\}$ and $\bm{y}^{(j)} = \{y_1, \ldots , y_j\}$ denote the sequence of allocations and responses observed for the first~$j$ patients (where both $\bm{a}^{(0)}$ and $\bm{y}^{(0)}$ are defined as the empty set). RAR defines the allocation probability $\pi_{k,i}$ conditional on $\bm{a}^{(i-1)}$ and $\bm{y}^{(i-1)}$, i.e.\ \begin{equation}
\label{eq:alloc_prob}
\pi_{k,i} = Pr\!\left(a_{k,i} = 1 \, | \, \bm{a}^{(i-1)}, \bm{y}^{(i-1)}\right).
\end{equation}
\thirdround{Note that for a procedure to be response-adaptive, the $\pi_{k,i}$ must depend on both $\bm{a}^{(i-1)}$ and $\bm{y}^{(i-1)}$}.
This framework is flexible enough to allow for the RAR procedure to \secondround{also} depend on covariates \secondround{that may affect the primary outcome}. Letting $\bm{x}^{(j)} = \{\bm{x}_1, \ldots , \bm{x}_j\}$ be a vector of observed covariates, we  define a \thirdround{C}ovariate-\thirdround{A}djusted \thirdround{R}esponse-\thirdround{A}daptive (CARA) procedure by letting $\pi_{k,i} = Pr\left(a_{k,i} = 1 \, | \, \bm{a}^{(i-1)}, \bm{y}^{(i-1)}, \bm{x}^{(i)}\right)$. 
With the increasing interest in ``precision  medicine'', the role of covariates is crucial in developing targeted therapies for patient subgroups.
\secondround{Many of the issues we discuss here for RAR are directly applicable (to some degree) to CARA. However, we do not include a specific discussion for CARA to preserve the focus of our work on RAR}. 
We instead refer the reader to the review by~\citet{Rosenberger2008}, more recent 
papers by~\citet{Atkinson2011, Antognini2011, Antognini2012, Metelkina2017} and the book by~\citet{Sverdlov2016}. 
\citet{Zagoraiou2017} 
discusses how to choose a CARA procedure in practice.
}

\changes{A final concept to introduce is that of hypothesis testing. We focus on the case where there is a global null hypothesis $\mathcal{H}_0: \theta_k = \theta_0 \, \forall k$ versus one-sided alternatives $\mathcal{H}_{1,k}: \theta_k > \theta_0$ for some~$k$ (assuming a larger value of $\theta_k$ represents a desirable outcome). At the end of the trial, a test statistic denoted $T_n = t(\bm{a}^{(n)}, \bm{y}^{(n)})$ is  computed based on the observed data. \secondround{The specific form of the test statistic depends on the outcome model and the hypothesis of interest. For example, if the  primary outcome is binary}, the \thirdround{M}aximum \thirdround{L}ikelihood \thirdround{E}stimator (MLE) of the success rate on treatment $k$ is
$\hat{p}_k = \frac{\sum_{i=1}^{n} a_{i,k}y_{i,k}}{\sum_{i=1}^{n} a_{i,k}}$. \secondround{In a two-arm trial}, one could use a $Z$-test based on the MLE of the success rates: \begin{equation}
\label{eq:Ztest}
Z_n = \frac{\hat{p}_1 - \hat{p}_0}{\sqrt{\hat{p}_0 (1 - \hat{p}_0)/N_0 + \hat{p}_1 (1 - \hat{p}_1)/N_1}}.
\end{equation}} \\[-36pt]


\subsection{\changes{Assessing the performance of RAR procedures} 
\label{subsec:performance}}


\changes{
In the literature, many ways of assessing RAR \secondround{have been considered}. Most metrics used in the clinical trial setting focus on \secondround{inferential} goals. 
Terms such as `power' and `patient benefit' can have very different meanings depending on the trial context. Here, rather than providing an exhaustive list of all possible metrics for comparing variants of RAR, we present some of the most relevant ones in three categories: \secondround{\emph{testing}}, \emph{estimation} and \emph{patient benefit}.} \\ 

\noindent \changes{\textit{\secondround{Testing} metrics: type~I error and power}}

\changes{
For confirmatory trials, the control of \secondround{frequentist} errors is especially important from a regulatory perspective. A type~I error is defined as falsely rejecting a null hypothesis~$\mathcal{H}_0$. 
For a trial with a single null hypothesis $\mathcal{H}_0: \theta = \theta_0$, the type~I error rate is defined as $\alpha = Pr(\text{rejecting } \mathcal{H}_0 \, | \, \theta = \theta_0)$, and for confirmatory trials this is controlled below some fixed level (typically $0.05$ or $0.025$). When there are multiple null hypotheses, various generalizations can be considered, the most common being the familywise error rate, which is the probability of making at least one type~I error. This reflects the inherent multiplicity problem and type~I error inflation that can occur if multiple hypotheses are tested without adjustment.
}

\changes{In contrast, a type~II error is failing to reject $\mathcal{H}_0$ when it is in fact false.
For a trial with a single null hypothesis $\mathcal{H}_0$ and corresponding point alternative hypothesis $\mathcal{H}_1: \theta = \theta_1$, the power of the trial is defined as $1-\beta = Pr(\text{rejecting } \mathcal{H}_0 \, | \, \theta = \theta_1)$. However, when there are multiple hypotheses (e.g. in 
the multi-arm setting with $K>1$), the `power' of the trial \secondround{admits} various definitions. For instance, marginal power \secondround{(the probability of rejecting a particular non-null hypothesis)}, disjunctive power \secondround{(the probability of rejecting at least one non-null hypothesis)} and conjunctive power \secondround{(the probability of rejecting all non-null hypotheses)} are all used as definitions of `power'~\citep{Vickerstaff2019}. 
\secondround{Additionally,} some authors define power as the probability of satisfying a criterion that reflects the goal of the trial. For example, power could be defined as the probability of selecting the best experimental treatment at the end of the trial, or as a Bayesian concept such as posterior predictive power. \thirdround{A RAR procedure can have a high power according to one definition but not according to another.}
} \\

\noindent \changes{\textit{Estimation metrics}}

\changes{There are metrics related to {estimation} \secondround{and the information gained} after a trial. A key consideration (particularly for adaptive designs, see~\citet{Robertson2021}) is bias, defined as a systematic tendency for the estimate of the treatment effect to deviate from its true value. More formally, the mean bias of an estimator $\hat{\theta}_k$ for $\theta_k$ is defined as $E(\hat{\theta}_k) - \theta_k$. An estimator may be biased due to the trial adaptations affecting its sampling distribution, or due to heterogeneity in the observed data (i.e.\ where the data does not come from the same underlying distribution, \secondround{such as when there is a time trend in the response variable as considered} in Section~\ref{subsec:time_trends}).
Apart from bias, another important consideration is the variance $\text{var}(\hat{\theta}_k)$ or mean squared error of an estimator $E [(\hat{\theta}_k - \theta_k)^2]$, reflecting the classical bias-variance trade-off. Although precision of the estimates is less often reported in the literature, this can be compared using estimation efficiency measures, see for example~\citet{Flournoy2013, Sverdlov2013a}.
} \\

\noindent \changes{\textit{Patient benefit metrics}}

\changes{Different metrics to capture the ``ethical" or {patient benefit} properties of RAR have been considered.
These are less frequently reported than \secondround{testing} and estimation metrics, which is somewhat counter-intuitive given the most common motivation to use RAR is to better treat more patients in a trial. Nevertheless, this lack of reporting is consistent with inferential goals being paramount. Some examples of patient benefit metrics include:
\begin{itemize}
    \item The number of treatment successes (for binary outcomes) or the total response (for continuous outcomes) in the trial: $\sum_{i=1}^n Y_i$. When averaged for binary outcomes, this is referred to as the \thirdround{E}xpected \thirdround{N}umber of \thirdround{S}uccesses (ENS). Alternatively, 
    some authors focus on the number of treatment failures $\sum_{i=1}^n (1 - Y_i)$ and report the \thirdround{E}xpected \thirdround{N}umber of \thirdround{F}ailures (ENF).
    \item The proportion of patients allocated to the best arm: $p^* =\sum_{i=1}^n a_{i,k^*}/n$, where $k^* = \text{argmax}_k \theta_k$ \secondround{(if $k^*$ is not unique then one option is to sum over all arms that are `best')}. 
\end{itemize}
%
\thirdround{The above metrics are concerned with the individual ethics of the~$n$ patients within the trial, which is distinct from the collective ethics of the overall population (which is related to testing and estimation metrics).}
We return to this issue of patient horizon 
in Section~\ref{subsec:ethics}. \\
}

\noindent \thirdround{\textit{Other metrics}}

\thirdround{Aside from the three categories of metrics described above, there are also metrics focusing on the level of imbalance in the number of patients in each treatment arm at the end of the trial. \secondround{One way of defining the imbalance in arm~$k$ is $(N_k/n - 1/(K+1))$}, which makes a comparison between the observed allocation ratio and a completely balanced allocation between the arms. See also Section~\ref{subsec:imbalance} \secondround{for other examples of imbalance metrics.}}

\changes{
A final metric is the total \emph{sample size} of the trial. Typically, this is defined as the minimum number of patients required to achieve a target power (given type~I error constraints) under some pre-specified point alternative hypothesis. This is closely linked with \secondround{testing metrics} but there are patient benefit considerations as well. For example, suppose one out of the~$(K+1)$ treatment options is substantially better than the rest. Using ER means that $Kn/(K+1)$ of the patients within the trial will be allocated to suboptimal treatments. Hence, minimizing the sample size~$n$ has patient benefit advantages as well. In contrast (as discussed in \citep{Berry1995}), increasing the sample size to maintain power when using RAR may deliver higher overall patient benefit across the target population (i.e.\ including future patients), suggesting trade-offs between benefit for patients in the trial and those outside of it\thirdround{, see also Section~\ref{subsec:ethics}}.}\\[-6pt] 
\section{A historical perspective on RAR} 
\label{sec:history}


\begin{displayquote}
``Those who cannot learn from history are doomed to repeat it.'' (Attributed to George Santayana)
\end{displayquote}


\noindent \changes{We now give an overview of the historical development of RAR, which naturally motivates how we classify RAR procedures in Section~\ref{subsec:taxonomies}. A \secondround{distinguishing feature of \thirdround{this} history is that a large amount of high quality theoretical work 
is paired with few highly influential examples} \changes{of RAR in practice. We thus present \thirdround{the} history \thirdround{of RAR}} in two distinct areas: theory (Section~\ref{subsec:methodology}) and practice} (Section~\ref{subsec:clinical_practice}).   
\changes{A timeline summarizing \thirdround{some} key developments is given in Figure~\ref{fig:timeline}.} \\

\definecolor{arrowcolor}{RGB}{144,168,65}
\colorlet{circlecolor}{white}
\definecolor{bordercolor}{RGB}{168,89,65}
\colorlet{textcolor}{bordercolor}
\setlength\taskwidth{2cm}

\begin{figure}[ht!]

\resizebox{\textwidth}{!}{
    \begin{timeline}
	\Task[1933]{Thompson sampling}
	\Task[1978]{RPW rule}
    \Task[1985]{ECMO trial}
    \Task[2000]{J\&T optimization approach}
    \Task[2001]{RSIHR paper}
    \Task[2004]{DBCD}
    \Task[2008]{BATTLE trial}
    \Task[2009]{ERADE}
    \Task[2010]{I-SPY 2 trial}
    \Task[2015]{Non-myopic RAR}
    \Task[2020]{REMAP-CAP trial}
\end{timeline}
}

    \caption{Timeline summarizing some of the key developments around the \secondround{theory and practice} of RAR in clinical trials. J\&T = \citet{Jennison2000}, RSIHR = \citet{Rosenberger2001}. \label{fig:timeline}}

\end{figure}



\subsection{RAR methodology}
\label{subsec:methodology}

The origins of \changes{RAR} \secondround{date back to~}\citet{Thompson1933}, who suggested allocating patients to the more \changes{promising} treatment arm via a posterior probability computed using interim data. \changes{RAR seems to have been the first form of an \emph{adaptive design} ever proposed.} 
\changes{Another influential early procedure was} the play-the-winner rule, proposed by \citet{Robbins1952} and then \citet{Zelen1969}. \changes{Although partially motivated by Thompson's idea, this is a non-randomized (deterministic) rule, where} a success on one treatment leads to \thirdround{the} subsequent patient being assigned to that treatment, while a failure leads to \thirdround{the} subsequent patient being assigned to \thirdround{the} other treatment. 

RAR also has roots in the methodology for sequential stopping problems (where the sample size is random), as well as bandit problems (where resources are allocated to maximize the expected reward). Since most \changes{of the earlier} work in these areas \changes{is} non-randomized (i.e.\ \thirdround{concerns} deterministic \changes{solutions}), we do not \changes{review them} here. 
\citet[Section 10.2]{Rosenberger2016} gives a brief summary of the history of both of these areas, and an overview of multi-arm bandit models is presented in~\citet{Villar2015a}. For a review of non-randomized algorithms for the two-arm bandit problem, see~\citet{Jacko2019}. 

\secondround{An important development for the clinical trials setting was the introduction of randomization to otherwise deterministic response-adaptive procedures. Randomization is essential for mitigating biases and ensuring comparability of treatment groups and is the default patient allocation mode in \thirdround{confirmatory} clinical trials~\citep{Rosenberger2016}. }
An \changes{example of this is} the \thirdround{R}andomized \thirdround{P}lay-the-\thirdround{W}inner (RPW) rule proposed by \citet{Wei1978}. The RPW rule can be viewed as an \textit{urn model}: each treatment \changes{allocation} is made by drawing a ball from an urn (with replacement) and the composition of the urn is updated based on the responses. In the following decades, many RAR rules based on urn models were proposed, with a focus on generalizing the RPW rule. We refer to~\citet[Chapter~4]{Hu2006} and~\citet[Section~10.5]{Rosenberger2016} for a detailed description. 


Urn-based RAR procedures are intuitive, but are not optimal \changes{designs} in \changes{a} formal \changes{mathematical} sense \changes{(see Section~\ref{subsec:taxonomies})}. \changes{From} the early 2000s a 
perspective on RAR emerged based on \textit{optimal allocation targets}, which are derived as a solution to a formal optimization problem. For two-arm \changes{group sequential} trials, a general optimization approach was proposed by~\citet{Jennison2000, Jennison2001}, \changes{which minimizes the expected value of a loss function which is an arbitrary weighted average of~$N_0$ and~$N_1$. This} led to the development of a whole class of optimal RAR designs. \secondround{An early example for \changes{two-arm} trials with binary outcomes is~\citet{Rosenberger2001}}. More examples 
\secondround{are given} in Section~\ref{subsec:power}. In order to \changes{implement} optimal allocation targets, a key development was the modification by~\citet{Hu2004a} of the \secondround{Doubly-adaptive Biased Coin Design (DBCD)} originally described by~\citet{Eisele1994}. Subsequent theoretical work by~\citet{Hu2006} focused on asymptotically best RAR procedures, \changes{i.e.\ those with minimum asymptotic variance of the \thirdround{optimal} allocation ratio} \thirdround{(which typically depends on unknown parameters that need to be estimated using the response data, see the equations in Section~\ref{subsec:power}). This} led to the development of the class of efficient RAR designs (\secondround{known as} ERADE) proposed by~\citet{Hu2009}.

All the RAR procedures above are \textit{myopic}, in that they use past responses \changes{$Y_{k,i}$ and past allocations $a_{k,i}$} to determine the  allocation probabilities $\pi_{k,i}$, without considering future patients \thirdround{to be recruited into the trial} and the information they could provide. A recent 
development is 
non-myopic or \textit{forward-looking} RAR  based on solutions to the multi-bandit problem. The first such 
procedure was 
by~\citet{Villar2015b} for 
binary responses, with subsequent work by~\citet{Williamson2017} accounting for a finite time-horizon 
and for normally-distributed outcomes (\thirdround{see}~\citet{Williamson2020}). \\

\subsection{RAR in clinical practice}
\label{subsec:clinical_practice}

\changes{One of} the earliest uses of RAR in clinical practice was the
ECMO trial~\citep{Bartlett1985}. This trial used the RPW rule on a study of critically ill babies randomized either to ECMO or to the conventional treatment. In total, 12~patients were observed: one in the control group, who died, and~11 in the ECMO group, who all survived. This extreme imbalance in sample sizes was a motivation for running a second randomized ECMO trial, using fixed randomization~\citep{Ware1989}.

\changes{These} ECMO trials have been the focus of much debate, with these two papers accruing over 1000 citations. Indeed, to this day the first ECMO trial is regarded as \secondround{a key reason not to} use RAR in clinical practice, due to the extreme treatment imbalance and highly controversial interpretation~\citep{Burton1997}. \changes{Most recently, \citet{Proschan2020} states ``[RAR] had an inauspicious debut in the aforementioned ECMO trial''.} 
\changes{Largely} due to the controversy around these trials, there was little use of RAR in clinical trials for the subsequent 20~years. \changes{The pace of methodological work on RAR and adaptive designs more generally was negatively impacted as well~\citep{Rosenberger2015}.} One exception was the Fluoxetine trial~\citep{Tamura1994}, which again used the RPW rule, but with a burn-in period to \secondround{avoid large imbalances in treatment groups.} \changes{For an in-depth discussion of both 
trials we refer to~\citet{Grieve2017}, which also discusses two BRAR trials from the early 2000s.}

\changes{More recently,} there have been high-profile clinical trials that use BRAR 
\changes{as a key (but not the only) part of their adaptive design.
Some} examples in oncology \changes{include} the BATTLE trials and the I-SPY~2 trial. The BATTLE
trials \citep{Kim2011, Papadimitrakopoulou2016} used RAR based on a Bayesian hierarchical model, where the randomization probabilities are proportional to the observed efficacy based on the individual biomarker profiles. Similarly, the I-SPY~2
trial~\citep{Barker2009} used RAR based on Bayesian posterior probabilities specific to different biomarker signatures. These trials have generated valuable discussions about the benefits and drawbacks of using RAR in clinical trials \citep{Das2017, Korn2017, Siu2017}. \changes{\secondround{Meanwhile}, the REMAP-CAP platform trial~\citep{Angus2020} also incorporated BRAR as part of its design, in the context of community-acquired pneumonia. This trial was subsequently tailored to respond to the COVID-19 pandemic~\citep{REMAP2021}.} 

\changes{Although the BATTLE, I-SPY~2 and REMAP-CAP trials use RAR as part of their designs, their primary focus was to select optimal treatments for particular biomarker signatures\thirdround{, and} \secondround{hence \thirdround{can more precisely be} described as master protocol trials~\citep{Woodcock2017}}. Arguably the main feature of I-SPY~2 was the mechanism to `graduate' or drop treatments and to add new ones as they arise. For recent examples of clinical trials using BRAR in a `vanilla' fashion (although still including early stopping rules), we refer to~\citet{Faseru2017, OBrien2019, Barohn2021}.} \\

\subsection{Classifying procedures: \changes{a taxonomy of RAR}}
\label{subsec:taxonomies}


\secondround{
Some papers (perhaps unintentionally) criticize the use of RAR in general or make broad conclusions using arguments that only apply to a specific class of procedures, as is (still) the case for the RPW rule and the ECMO trial \citep{Proschan2020}. In reality, RPW is just one example of a RAR procedure out of many and hence the value of other RAR procedures that are markedly different is harder to see.}
The vast number of different RAR procedures is a challenge that non-experts and experts alike face with when exploring the literature, which has accumulated \changes{(and continues to quickly evolve)}.

\secondround{We now define} several families of RAR procedures \changes{and} discuss how they fit different classification criteria. \changes{This discussion illustrates the wealth and breadth of RAR methodology and its importance when assessing its value} for a specific application. However, \changes{the} criteria \changes{are not} exhaustive \changes{or able to} completely differentiate all \changes{types of RAR}. \secondround{As discussed next, we expect most classifications to require frequent revisiting given the current pace of development in the area~\citep{Villar2020}}. \changes{Nevertheless, these classifications \secondround{can allow a better understanding of the many existing approaches and how they compare.}}
\changes{We note that the number of references of each RAR family throughout the paper is a reflection of the attention each method received in the past rather than an intended focus.}\\
%


\noindent \textit{Optimal and design-driven RAR}

An \changes{important broad distinction first described  by~\citet{Rosenberger2002, Hu2004a} is between `optimal' and `design-driven' RAR. In their works, this is defined as \secondround{the following}.} \\[-6pt]

\begin{enumerate}
    \item \changes{`\emph{optimal}' RAR}  is based on \changes{deriving} an optimal allocation target \changes{(or a sampling ratio), by optimizing a} specific criterion based on a population response model.\\[-6pt]
    
    E.g.\ \changes{In} \citet{Rosenberger2001} an \secondround{optimal RAR} is defined for a two-arm trial based on the population model \changes{for binary responses}.  The power \changes{at the end} of the trial (using a $Z$-test \changes{as given in equation~\eqref{eq:Ztest}}) is fixed, \changes{while} the \changes{ENF} is minimized. \changes{Formally, using the notation in Section~\ref{subsec:notation} and defining $\rho = N_1/n$, the optimization problem is as follows:
    \begin{equation}
    \label{eq:RSIHR}
    \min_{\rho} \{(1-p_0)N_0 + (1-p_1)N_1 \} \quad \text{subject to} \quad \frac{p_0 (1 - p_0)}{N_0} + \frac{p_1 (1 - p_1)}{N_1} = C \end{equation}
    The solution $\rho^*$ is then the optimal target ratio (given the above optimization criteria). To implement this in practice, it is necessary to estimate the parameters $p_0$ and $p_1$.} \\
    
    \item \changes{\emph{`design-driven'} RAR is based on} rules \changes{which} are established with intuitive motivation, but are not optimal in a formal sense.\\[-6pt]
    
    E.g.\ The RPW rule for binary responses.
    The rules for computing and choosing the allocation probability can be formulated using an intuitive urn-based model (see \changes{Section~\ref{subsec:methodology} and} \citet{Wei1978} for details). \\[-6pt]
\end{enumerate}






\secondround{A key difference for \changes{these two RAR classes} is the computation of allocation probabilities.}  \changes{While} approaches in family~(1) rely on optimizing \changes{some} objective function that describes aspects of the population model explicitly, those belonging to family~(2) typically have \changes{an} intuitive motivation that is not defined analytically from a population model.
\changes{However, while classifying approaches into these two families is useful, there are some important caveats. 
First, an intuitive design may eventually be formally shown optimal in some sense.} 
\secondround{Second, some procedures are harder to classify into the above criteria}. Consider bandit-based designs, \changes{such as the \thirdround{F}orward-\thirdround{L}ooking Gittins \thirdround{I}ndex (FLGI) rule  in~\citep{Villar2015b} or the design by~\citet{Williamson2017}.
These} are based on an optimization approach but do not explicitly target a pre-specified optimal allocation ratio like in family~(1). 
In certain cases (like for FLGI), these are heuristic approximations and can be viewed as having a more intuitive motivation. 

\secondround{An final caveat is that there are different optimality notions to consider. \textit{Asymptotic} optimality for example was first introduced by~\citet{Robbins1952}}. 
For example, \changes{TS} is asymptotically optimal \secondround{in terms of} minimizing cumulative regret (see e.g.~\citet{Kaufmann2012}). \changes{So for large trials, one could consider it as belonging to family~(1). However, in small} samples, if TS (and its generalization proposed by~\citet{Thall2007}) is \secondround{used for} assigning more patients to the better arm, then this would be closer an intuitive motivation \secondround{(as in family~2), as only dynamic programming achieves ENS optimality in a finite sample}.  \\


\noindent \changes{\textit{Parametric and non parametric RAR}}

A classification that follows naturally from the previous one is that of \emph{parametric} and \emph{non-parametric} (or \emph{distribution free}) RAR. This classification captures some of the spirit of the \emph{optimal} versus \emph{design-driven} while being possibly less subject to caveats. Parametric RAR procedures rely on assumptions that the response data are drawn from a given parametric probability distribution to compute and update the allocation probabilities $\pi_{k,i}$.

\begin{displayquote}
E.g.\ the optimal RAR procedure proposed in \citet{Rosenberger2001} and defined above requires estimates of $p_0$ and $p_1$ in order to determine $\pi_{k,i}$.
\end{displayquote}

In contrast, non-parametric RAR procedures do not explicitly rely on 
a parametric probability distribution nor on the corresponding parameter estimates to compute and update~$\pi_{k,i}$.

\begin{displayquote}
E.g.\ the RPW rule (and urn designs) are non-parametric designs that can be used for any binary data, regardless of the underlying probability distribution. \\
\end{displayquote}

\noindent \textit{Bayesian and frequentist \changes{RAR}}

\secondround{The distinction of RAR based on the frequentist or Bayesian approach to statistics may apply to the inference procedure used for the final analysis and/or to the design of the RAR itself.} 
In our opinion, the \changes{inferential classification} may not be helpful, 
since the choice of inference procedure depends on the experimental goals and regulators' preferences between these approaches. Moreover, some innovative approaches have Bayesian design aspects but the inference focuses on the frequentist operating characteristics, \changes{see e.g.\ \citet{Ventz2017}}. \changes{Arguably a more relevant element to consider is} the objective(s) of RAR \changes{(see the subsection `RAR with different objectives' below)}. Readers interested in understanding the pros and cons of frequentist and Bayesian inference are referred to 
\citet{Wagenmakers2008, Samaniego2010} as this is outside the scope of our review. \changes{For references on the use of Bayesian designs in the clinical trial context, we refer to~\citet{Chow2007, Chevret2012, Rosner2020, Stallard2020}}.

\changes{A common definition of a Bayesian design is that} a prior distribution \changes{is} explicitly incorporated into the design criteria/optimization problem and/or into the calculation of the allocation probabilities. \changes{However, the use of a prior distribution is not the defining element of BRAR as one can sometimes find equivalent {frequentist} designs using penalized MLEs or a specific prior distribution.} For example,  where the posterior mode \changes{with a uniform prior} coincides with the MLE in a RAR procedure then an update of probabilities is the same from a frequentist and Bayesian perspective (\changes{see also a hybrid formulation for the RPW rule given in~\citet[pg.~271]{Atkinson2014}).}

\changes{Hence, in the context of RAR, we define a Bayesian design as ``a design rule that depends recursively on the posterior probability of the parameters'' \citep{Atkinson2014}, where the recursive updating of the allocation probabilities is done via Bayes Theorem. \thirdround{The prior information itself can be updated at time points when accrued trial data is available, see~\citet{Sabo2014}}. Such designs are called ``fully Bayesian'' in~\citet{Ryan2016}, and allow the full probabilistic description of all uncertainties, including future outcomes (i.e.\ predictive probabilities).} 

\begin{displayquote}
E.g.\ In TS with \secondround{$K=1$} and binary responses, the randomization probability is the posterior probability that $p_1 > p_0$ (given the prior information and available trial data), i.e.\ $\pi_{1,i} = P(p_1 > p_0 \, | \, \bm{a}^{i-1}, \bm{y}^{i-1})$. 
\end{displayquote}

\noindent \changes{A RAR} procedure is \emph{frequentist} if a frequentist approach is used for \changes{both} estimating the unknown parameter(s) \changes{and, more importantly, for updating the allocation probabilities}.  

\begin{displayquote}
E.g.\ \changes{the DBCD can be used to target different allocations (see Section~\ref{subsec:power}), where the $\pi_{k,i}$ are given as functions of the MLE.} \\
\end{displayquote}

\noindent \changes{\textit{RAR methodological families}}

 RAR procedures can be classified in terms of the broad methodological `families' they belong to: \changes{\ RAR based on TS (e.g. those suggested by~\citet{Thall2007}), RAR based on urn models} \changes{(e.g. RPW)}, RAR that target a pre-specified (optimal) allocation ratio \changes{(e.g. as in~\citet{Hu2004a})} or bandit-based RAR procedures \changes{(e.g., the FLGI)}. \changes{This classification naturally follows from the historical developments in the area.} However, 
 RAR procedures could conceivably belong to more than one family and new types of RAR are continuously being developed.\\

\noindent\changes{ \textit{RAR with different objectives}}

 RAR procedures differ in the goal they are designed to achieve, \changes{either formally or intuitively}. While some  consider competing objectives such as both \changes{power} and patient benefit (see Section~\ref{subsec:performance} for definitions), others prioritize one over the other. Additionally, procedures can be non-myopic or myopic \changes{in their objective formulation}. For some RAR procedures, such as those targeting an optimal allocation, the optimization problem can account for multiple objectives, see e.g.\ \citet{Hu2015, Antognini2010, Antognini2015}. \changes{More generally, within a Bayesian framework there is scope for composite utilities for multi-objective experiments~\citep{McGree2012, Antognini2015, Metelkina2017}}. 
 \secondround{The selection of an objective may also \changes{require computational considerations}}.

\changes{Therefore, a good classification for comparing performance of RAR procedures 
is that of \emph{single} objective procedures versus those that have \emph{composite} objectives (reflecting trade-offs and constraints between possibly competing goals of an experiment).}

\begin{displayquote}
\changes{
E.g.\ FLGI in~\citet{Villar2015b} has a non-myopic patient benefit goal, while Neyman allocation (see Section~\ref{subsec:power}) has a power goal.}
\end{displayquote}

\begin{displayquote}
\changes{
E.g.\ the design in~\cite{Williamson2017} has a non-myopic patient benefit goal subject to a power constraint, while the `optimal' allocation of~\citet{Rosenberger2001} has a myopic patient benefit goal also subject to a power constraint.} \\
\end{displayquote}

\section{\protect\changes{Established views} on RAR}
\label{sec:myths}


\secondround{In this section, we 
critically examine some \changes{published views} on RAR. \changes{We present them labeled as questions because we have received them as such during informal exchanges with trial statisticians.} }
\changes{We} 
provide a \secondround{complementary} view of the use of RAR procedures, which acknowledges problems and disadvantages, but also emphasizes the solutions and advantages.

In what follows, we \changes{base our discussion on} specific examples of RAR procedures \changes{only as a way} to illustrate \changes{how some established views} on RAR do not hold in general. The examples used below are by no means presented as the `best' RAR procedures, or even necessarily recommended for use in practice -- such judgments critically depend on the context and goals of the specific trial under consideration. 
\changes{We direct the reader to Section~\ref{sec:discuss} for 
the latter point}.
\\


\subsection{Does RAR lead to a substantial chance of allocating more patients to an inferior treatment? }

\label{subsec:imbalance}

\citet{Thall2016} give a number of undesirable properties of RAR, including the following:

\begin{displayquote}
\ldots there may be a surprisingly high probability of a sample size imbalance in the wrong direction, with a much larger number of patients assigned to the inferior treatment arm, so that \changes{[RAR]} has an effect that is the opposite of what was intended.
\end{displayquote}

\noindent
\secondround{In simulation studies of two-arm trials with a binary outcome in~\citet{Thall2015, Thall2016} TS is shown to have a substantial chance (up to 43\% for the parameter values considered) of producing sample size imbalances in the wrong direction (i.e.\ the inferior arm) of more than 20 patients out of a maximum of 200. }
While this result \changes{holds for the specific BRAR procedure in the scenarios under consideration in that work,} 
these conclusions \secondround{do} not hold for \secondround{all} types of RAR. \changes{These authors} were among the first to compute this 
metric of sample size imbalance, and most of the RAR literature \changes{does} not report it (or related ones). Hence it is unclear how other families of RAR procedures perform \secondround{in this} regard.
To \changes{address} this, we perform a new simulation study 
in the two-arm trial setting with a binary outcome.
We compare the \secondround{following range} of RAR procedures:

\begin{itemize}
\setlength\itemsep{6pt}

    
    \item \textit{Permuted block randomization [PBR]}: patients are randomized in blocks to the treatments so \thirdround{that} exact balance is achieved for 
    each block (and hence at the end of the trial).
    
    \item \textit{Thall and Wathen [TW(c)]}: randomizes patient~$i$ to 
    treatment $k=1$ with probability \changes{\[
    \pi_{1,i} = \frac{\left[P(p_1 > p_0 \, | \, \bm{a}^{i-1}, \bm{Y}^{i-1})\right]^c}{\left[P(p_1 > p_0 \, | \, \bm{a}^{i-1}, \bm{Y}^{i-1})\right]^c + \left[1 - P(p_1>p_0 \, | \, \bm{a}^{i-1}, \bm{Y}^{i-1})\right]^c}
    \]}
    
    Here \changes{$P(p_1>p_0 \, | \, \bm{a}^{i-1}, \bm{Y}^{i-1})$} is the posterior probability that \changes{the experimental treatment has a higher success \thirdround{rate} than the control treatment}. The parameter~$c$ controls the variability of the procedure. Setting~$c = 0$ gives \changes{ER}, while setting~$c=1$ gives TS \changes{as described in Section~\ref{subsec:taxonomies}}. \citet{Thall2007} suggest setting~$c$ equal to~$1/2$ or~$i/(2n)$. 

    \item \textit{Randomized Play-the-Winner Rule [\thirdround{RPW}]}: 
    see \changes{Section~\ref{subsec:methodology} and}~\citet{Wei1978}. 
    
    \item \textit{Drop-The-Loser rule [DTL]}: 
    \secondround{a generalization of the RPW} proposed by~\citet{Ivanova2003}. 

    \item \textit{Doubly-\secondround{adaptive} Biased Coin Design [DBCD]}: a response-adaptive procedure targeting the optimal 
    ratio of~\citet{Rosenberger2001}. 
    For details, see~\citet{Hu2004a}. 
    
    \item \textit{Efficient Response-Adaptive Randomization Designs [ERADE]}: a response-adaptive procedure targeting the optimal allocation ratio of~\citet{Rosenberger2001}. It attains the lower bound of the allocation variances, see~\citet{Hu2009} for further details. 
    
    \item \textit{Forward-looking Gittins Index [FLGI\changes{(b)]}}: a 
    \secondround{ 
     RAR procedure with near-optimal patient benefit properties proposed in~\citet{Villar2015b}}. This depends on a block size~$b$. 
    
    \item \textit{Oracle}: hypothetical \changes{non-randomized} rule that assigns all patients to the \textit{\thirdround{true}} best-performing arm \changes{(i.e.\ $\pi_{k^*, i}=1$ for $k^*=\max_k {p_k}$ and $\pi_{k,i}=0$ otherwise for all~$i$)}. 
    \\[-6pt]
    
\end{itemize}
 
In our simulations, we \changes{initially} set $p_0 = 0.25$ and vary the values of $p_1$ \thirdround{(with $p_1 > p_0$)} and~$n$. Unlike in~\citet{Thall2016}, we do not include early stopping in order to isolate the effects of using RAR procedures. We evaluate performance in terms of \changes{several imbalance metrics including} \secondround{$E(N_1-N_0)$ and the} (2.5 percentile, 97.5 percentile) of $(N_1 - N_0)$; the probability of a imbalance of more than 10\% of the total sample size in the wrong direction \secondround{(i.e.\ allocating more patients to the inferior arm)}, \changes{denoted} $\hat{S}_{0.1} = \text{Pr}(\secondround{N_0 > N_1} + 0.1n)$ when \secondround{$p_1 > p_0$};
the ENS and its standard deviation. Note that our measure of $\hat{S}_{0.1}$ coincides with the \changes{single imbalance} measure used in~\citet{Thall2016} when $n = 200$. 

Table~\ref{tab:imbalance} shows the results for $p_1 = 0.35$ and $n \in \{200, 654\}$. When $n = 200$, TS has a substantial probability \changes{($\hat{S}_{0.1} \approx$ 14\%)} of an \secondround{undesirable} imbalance in the wrong direction, while using the Thall and Wathen (TW) procedure reduces this probability, which (as expected) agrees with~\citet{Thall2016}. Unsurprisingly, the 
bandit-based procedures (\thirdround{i.e.}~FLGI) also has relatively large values of $\hat{S}_{0.1}$, although \changes{interestingly} these are still \changes{smaller} than for TS \changes{which could be due to their non-myopic nature. \secondround{Meanwhile,}} ER has $\hat{S}_{0.1} \approx 0.07$, which \secondround{provides a simple theoretical baseline (although in practice, for larger trials a form of PBR would be most suitable).}
In contrast, the RPW, DBCD, ERADE and DTL procedures all have values of $\hat{S}_{0.1}$ of $0.01$ or less, which is also reflected in the \secondround{ranges} for the sample size imbalance. \secondround{These procedures are hence comparable to PBR in terms of this imbalance metric.}

\begin{table}[ht!]
\caption{Properties of various patient allocation procedures, where $p_0 = 0.25$ and $p_1 = 0.35$. Results are from $10^4$ trial replicates. \label{tab:imbalance} \vspace{6pt}}

\centering
\setlength{\tabcolsep}{12pt}
\renewcommand{\arraystretch}{1.5}

\begin{tabular}{l l l l l}
 \hline
 $\bm{n}$ & \textbf{Procedure} & $\bm{N_1 - N_0}$  & $\bm{\hat{S}_{0.1}}$ & \textbf{ENS} \\ \hline
 200  & ER & 0 (-28, 28) & 0.069 & 60 (6.4) \\
 (Low power)& PBR & 0 & 0 & 60 (6.4) \\
 & Oracle & 200 & 0 & 70 (6.7) \\
 & TS & 95 (-182, 190) & 0.137 & 65 (8.5) \\
 & FLGI($b=5$) & 114 (-176, 190) & 0.111 & 66 (8.3) \\
 & FLGI($b=10$) & 115 (-172, 190) & 0.100 & 66 (8.2) \\
 & TW(1/2) & 74 (-90, 174) & 0.085 & 64 (7.5) \\
 & TW($i/2n$) & 50 (-28, 122) & 0.038 & 63 (6.8) \\
 & RPW & 14 (-16, 44) & 0.011 & 61 (6.5) \\
 & DBCD & 17 (-10, 46) & 0.003 & 61 (6.4) \\
 & ERADE & 16 (-6, 42) & 0.000 & 61 (6.4) \\ 
 & DTL & 14 (-4, 32) & 0.000 & 61 (6.6) \\ \hline
  654  & ER & 0 (-50, 50) & 0.005 & 196 (11.7) \\
 (High power)& PBR & 0 & 0 & 196 (11.6) \\
 & Oracle & 654 & 0 & 229 (12.2) \\
 & TS & 461 (-356, 640) & 0.042 & 220 (17.0) \\ 
 & FLGI($b=5$) & 511 (-619, 645) & 0.054 & 222 (18.5) \\
 & FLGI($b=10$) & 511 (-617, 645) & 0.051 & 222 (18.0) \\
 & TW(1/2) & 384 (44, 594) & 0.011 & 215 (14.2) \\
 & TW($i/2n$) & 272 (54, 456) & 0.010 & 210 (13.0) \\
 & RPW & 46 (-8, 100) & 0.000 & 199 (11.8) \\
 & DBCD & 55 (8, 106) & 0.000 & 199 (11.8) \\
 & ERADE & 54 (16, 96) & 0.000 & 199 (11.7) \\
 & DTL & 46 (14, 80) & 0.000 & 198 (11.7) \\ \hline
\end{tabular}

\end{table}

\secondround{The total sample size \changes{(in comparison to the treatment effect)} can have a large impact on these imbalance metrics.
When} $n = 200$, the trial has low power to declare \changes{the experimental treatment superior to the control}. If the sample size is chosen so that \changes{ER} yields a power of \changes{at least}  80\% (when using the standard $Z$-test), then \thirdround{we require} $n \geq 654$. For $n = 654$, Table~\ref{tab:imbalance} shows that the values of $\hat{S}_{0.1}$ are substantially reduced for TS, the TW procedure and the bandit-based procedures. The \secondround{ranges} for $N_1-N_0$ suggest that TS and the bandit-based procedures still have a small risk of getting `stuck' on the wrong treatment. 

\begin{figure}[ht!]
    \centering
    \includegraphics[width = 0.9\textwidth]{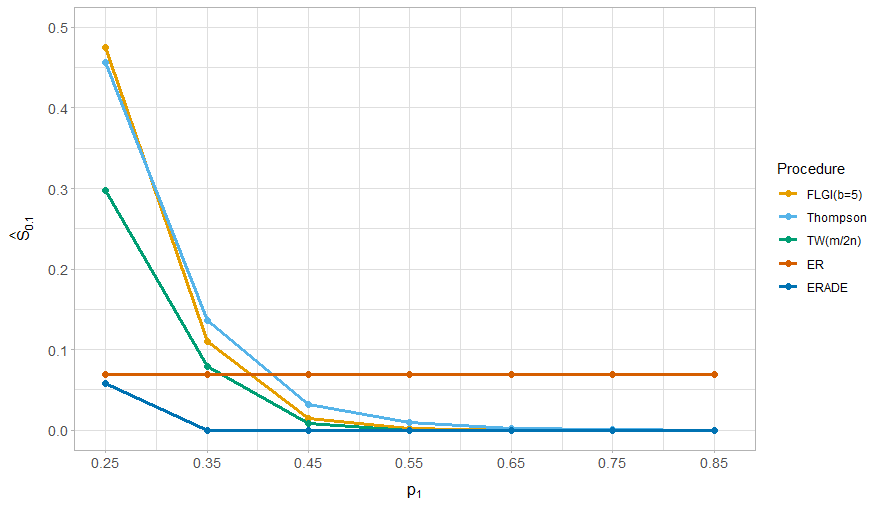}
    \caption{Plot of $\hat{S}_{0.1}$ for various RAR procedures as a function of $p_1$, where $p_0 = 0.25$ and $n = 200$. Each data point is the mean of $10^4$ trial replicates.}
    \label{fig:imbalance}
\end{figure}

Another important factor is the magnitude of the difference between~$p_0$ and~$p_1$ \changes{or the treatment effect}. The scenario considered above with $p_0 = 0.25$ and $p_1 = 0.35$ is a relatively small difference (as shown by the large sample size required to achieve a power of 80\%), and the more \changes{patient-benefit oriented} rules would not perform well in terms of sample size imbalance in this case. Table~\ref{tab:imbalance2} \changes{(in the Appendix)} shows the results when $p_1 = 0.45$ and $n = 200$. The values of $\hat{S}_{0.1}$ are substantially reduced for TS as well as for the TW and bandit-based procedures, being much less than for \changes{ER} \secondround{and not substantially greater than using PBR}. In terms of the mean and \secondround{ranges} for $N_1 - N_0$, these are now especially appealing for FLGI.
Figure~\ref{fig:imbalance} extends \changes{this} analysis by considering the value of $\hat{S}_{0.1}$ for a range of values of $p_1$ from 0.25 to 0.85 \changes{when $n=200$ to illustrate how this issue evolves as we move away from the null hypothesis scenario} \secondround{(while recognising that small differences of $p_1$ from $p_0$ may not be practically important)}. For $p_1$ greater than about 0.4, the probability of a substantial imbalance in the wrong direction is higher for a simple ER design than for all of the other RAR procedures considered.

Figure~\ref{fig:imbalance} demonstrates another \changes{issue of} $\hat{S}_{0.1}$ as a performance measure. This probability of imbalance increases for the RAR procedures considered as the difference $p_1 - p_0$ decreases, but as this difference decreases, so do the consequences of assigning patients to the \changes{inferior} treatment.  Table~\ref{tab:imbalance} depicts trade-offs between sample size imbalance (as measured by $N_1-N_0$ and $\hat{S}_{0.1}$) and the ENS. The most \changes{patient-benefit oriented} RAR procedures (TS and FLGI) have the highest ENS, which are in fact close to the highest possible ENS (the `Oracle' procedure). However, these procedures also perform the worst in terms of sample size imbalance. This demonstrates our general point that careful consideration is needed by looking at a variety of performance measures \secondround{instead of} focusing on a single measure such as $\hat{S}_{0.1}$. \\


\noindent \textit{Summary}

In summary, RAR procedures do not necessarily have a high probability of a substantial sample size imbalance in the wrong direction, when compared with using \changes{ER} \secondround{or PBR}. This probability crucially depends on the true \changes{treatment effect}, as well as the \changes{planned} sample size of the trial. \changes{These results suggest that sample size imbalance may be larger when the effect size is smaller (i.e.\ being close to the null), and we hypothesize that this may generalize beyond the binary context.} 

\secondround{If sample size imbalance is of particular concern in a specific trial context, \secondround{an option} is to consider the use of constraints to avoid imbalance, such as the constrained optimization approach of~\citet{Williamson2017}. Recently \citet{Lee2021} also  proposed an adaptive clip method (i.e.\ having a lower bound on the allocation probabilities) that can be used in conjunction with BRAR to reduce the chance of imbalance. Potential sample size imbalances need to be carefully evaluated in light of other performance metrics: restricting imbalance limits the potential for the patient benefit gains RAR can attain. Of course, if sample size imbalance needs to be strictly controlled in a trial, a restricted randomization scheme (such as \thirdround{PBR}) may be more appropriate than using RAR.} \\



\subsection{Does the use of RAR reduce statistical power?}
\label{subsec:power}

\changes{Perhaps} one of the \changes{most well established views} about RAR procedures is that their use reduces statistical power, as stated in \citet{Thall2015}:

\begin{displayquote}
Compared with an ER design, [RAR] \ldots [has] smaller power to detect treatment differences.
\end{displayquote}


\noindent Similar statements \secondround{appear} in \citet{Korn2011a} and \citet{Thall2016}. Through simulation studies, these papers \changes{(all focused on the two-arm setting \secondround{with binary outcomes})} show that \changes{ER} can have a higher power than BRAR \secondround{for a fixed sample size}, or equivalently that a larger sample size is needed for BRAR to achieve the same power and type~I error rate as an \changes{ER} design. 

These papers only consider the BRAR procedure proposed by~\citet{Thall2007} (see Section~\ref{subsec:imbalance} for a formal definition). 
%
%
%
\changes{As shown in~\citet{Hu2006}, RAR procedures will have additional variability introduced by the correlation between the outcome $Y_{k,i}$ and allocation $a_{k,i}$, and this will in turn translate into a higher variability $\text{var}(T_n)$ of a statistical test $T_n$ (hence reducing power). Yet, as we discuss, \secondround{there exist} RAR procedures that control for this, so that their use does not necessarily reduce power.}
\changes{In this section}, we focus solely on power considerations and we assume the use of standard \secondround{(frequentist)} inferential tests to make power comparisons, which we return to in Section~\ref{subsec:stat_inference}. \changes{Finally, we present the two-arm and multi-arm trial settings in distinct subsections below, since (as discussed in Section~\ref{subsec:performance}) the definition of `power' becomes more complex in the latter setting.} \\


\noindent \textit{Two-arm trials}

Some RAR procedures formally target optimality criteria as a reflection of the trial's objectives\thirdround{,} including power. \secondround{In a 
binary outcome setting}, as in~\citet{Rosenberger2004} with 
\thirdround{the} $Z$-test \changes{\thirdround{given} in equation~\eqref{eq:Ztest}} \secondround{and \thirdround{defining} $\rho=N_1/n$},
one strategy is to fix the power of the trial and find $(N_0, N_1)$ to minimize the total sample size~$n$.  \secondround{This is equivalent to} fixing~$n$ and finding $(N_0, N_1)$ to maximize the power. This gives the \changes{optimal} ratio known as Neyman allocation\thirdround{,}  $\rho_{\text{Neyman}}^*$:
\begin{equation}
\rho_{\text{Neyman}}^* = \frac{\sqrt{p_1 (1-p_1)}}{\sqrt{p_0 (1-p_0)} + \sqrt{p_1(1-p_1)}}\thirdround{.}
\end{equation}
\secondround{In general, $\rho_{\text{Neyman}}^* \neq 1/2$ and hence} \changes{ER} \changes{does not} maximize the power for a given $n$ when responses are binary. 
The notion that ER maximises power in general is \changes{an established} belief that appears in many papers (see e.g.\ \citet{Torgerson2000}) but it only 
%
%
%
\secondround{holds in specific settings (e.g. if comparing means of two normally-distributed outcomes with a common variance)}.

An \secondround{ethical problem with this} allocation \secondround{maximising power} is that if $p_0 + p_1 > 1$, more patients will be assigned to the treatment with the smaller $p_k$. This shows the potential trade-off between power and patient benefit and
%
%
\secondround{motivated} the alternative 
approach by \citet{Rosenberger2001} as in Section~\ref{subsec:taxonomies} \changes{-- see equation~\eqref{eq:RSIHR}}. The \changes{optimal} solution \changes{$\rho_{R}^*$ is as follows:}
\begin{equation}
\rho_{R}^* = \frac{\sqrt{p_1}}{\sqrt{p_0} + \sqrt{p_1}}.
\end{equation}

\begin{figure}[ht!]
    \centering
    \includegraphics[width = \textwidth]{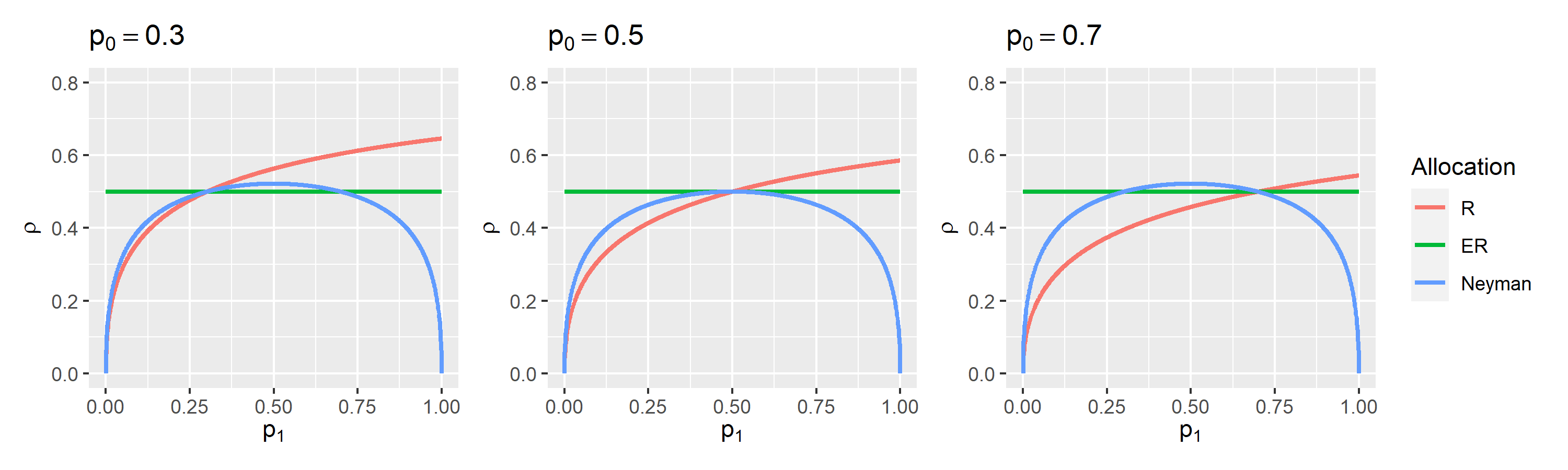}
    \caption{Plot of the optimal allocation ratios $\rho_{\text{Neyman}}^*$ and $\rho_R^*$ as a function of $p_1$, for $p_0 \in \{0.3, 0.5, 0.7\}$.}
    \label{fig:opt_alloc}
\end{figure}

\changes{Figure~\ref{fig:opt_alloc} shows the optimal allocation ratios $\rho_{\text{Neyman}}^*$ and $\rho_R^*$ as a function of $p_1$ for different values of $p_0$. Both coincide with ER only when $p_1 = p_0$ while 
$\rho_R^*$  always allocates more patients to the treatment which has the higher \thirdround{success rate}. Looking at $\rho_{\text{Neyman}}^*$, for $p_1 + p_0 < 1$ \secondround{a higher allocation to the treatment with the higher success rate will be more powerful than~ER}.} 

For \secondround{many types of endpoints, such as binomial and survival outcomes}, the model parameters in the optimization problem are unknown and need to be estimated from the accrued data. These estimates can then be used (for example)  \secondround{with} DBCD~\citep{Hu2004a} or ERADE~\citep{Hu2009} to target the optimal allocation ratio. Using the DBCD in this manner, \citet{Rosenberger2004} found in their simulation studies that it was 

\begin{displayquote}
\ldots as powerful or slightly more powerful than complete randomization in every case and expected treatment failures were always less
\end{displayquote}

\changes{Similar theoretical results are in~\citet{Yuan2011}}. This is consistent with a general guidelines given by~\citet{Hu2006} for using RAR procedures in a clinical trial, one of which is that \textit{power should be preserved}. 
%
RAR procedures that achieve this aim have been derived (in a similar spirit to the optimal allocation above) for continuous~\citep{Zhang2006} and survival~\citep{Zhang2007a} outcomes. \changes{Another line of work by~\citet{Antognini2018a, Antognini2018b} has looked at modifying the classical Wald test statistic for normally distributed outcomes in order to simultaneously improve power and patient benefit.}\\

\noindent \textit{Multi-arm trials}

\secondround{Similar concerns about `power'} 
for multi-arm RAR procedures \secondround{have been discussed}.  For example, \citet{Wathen2017} simulate a variety of five-arm trial scenarios and conclude

\begin{displayquote}
In multi-arm trials, compared to ER, 
several commonly used adaptive randomization methods give much lower probability of selecting superior treatments.
\end{displayquote}

\noindent Similarly, \citet{Korn2011b} simulate a four-arm trial and find that a larger average sample size is needed when using a RAR procedure instead of ER in order to achieve the same \changes{marginal} power. 
\secondround{As discussed in Section~\ref{subsec:performance}, there are different power definitions in this case}.
\citet{Lee2012} reach similar conclusions in the three-arm setting for disjunctive power. However, all these papers only consider variants of the TW procedure (the ``commonly used adaptive randomization methods'' quoted above) for multi-arm trials, and these conclusions \secondround{may not hold for RAR procedures in general}.


The optimal allocation  \changes{in~\citet{Rosenberger2001}} 
can be generalized for multi-arm trials, \changes{assuming a global null hypothesis}. The allocation is optimal in that it fixes the power 
\changes{to reject the global null} and minimizes the ENF. This was first derived by~\citet{Tymofyeyev2007}, who showed through simulation that for three treatment arms, using the DBCD to target the optimal allocation

\begin{displayquote}
\ldots provides increases in power along the lines of 2--4\% [in absolute terms]. The increase in power contradicts the conclusions of other authors who have explored other randomization procedures [for two-arm trials]
\end{displayquote}

Similar conclusions are given in~\citet{Jeon2010}, \citet{Sverdlov2013a} and~\citet{Bello2016}.

These optimal allocation procedures maintain (or increase) the power of the test \changes{to reject the global null}, but may have low marginal powers compared with \secondround{ER} 
in some scenarios, as shown in~\citet{Villar2015b}. However, even considering the marginal power to reject the null hypothesis \changes{$\mathcal{H}_{0,k^*}: \theta_{k^*} = \theta_0$} for the best treatment~$k^*$, \citet{Villar2015b} propose non-myopic RAR procedures (i.e.\ \thirdround{the ``controlled''} FLGI rules) that in some scenarios have both a higher marginal power and a higher ENS 
when compared with ER with the same sample size. 

Finally, the power comparisons made throughout this section have been against \changes{ER}. 
\secondround{A different comparison would be against} group-sequential and \thirdround{M}ulti-\thirdround{A}rm \thirdround{M}ulti-\thirdround{S}tage (MAMS) designs \changes{using ER} 
in each stage.
Both~\citet{Wason2014} and~\citet{Lin2017} show that BRAR can have a higher power than MAMS designs when there is a single effective treatment. \changes{More recently}, \citet{Viele2020a} show that the control allocation plays a part in achieving the power of a study when a variant of the TW procedure is implemented. These \changes{authors} also explore other design aspects in conjunction with the control allocation, and find that RAR can have \secondround{acceptable} power in some settings~\citep{Viele2020b}.\\

\noindent \textit{Summary}


In conclusion, if RAR is used to \secondround{improve patient benefit properties (in terms of ENF or ENS)}, 
then the power compared to \changes{ER} can be preserved  through an appropriate choice of the RAR procedure for the trial setting. Of course, this needs to be made with the objectives of the trial in mind (\changes{see Section~\ref{sec:discuss})}. 
If maximizing power is a key objective, then using \changes{ER} (instead of RAR) may not necessarily achieve this, \changes{even for two-arm trials}. \secondround{As discussed above, the nature of the response distribution plays an important role in these considerations, with much of the RAR literature focusing on binary responses}.\\


\subsection{\changes{Does 
RAR make valid statistical inference (more) challenging?}}
\label{subsec:stat_inference}

\secondround{The Bayesian approach to statistical inference allows the seamless analysis of results of a trial that uses RAR. However, as noted in~\citet{Proschan2020},} 

\begin{displayquote}
The frequentist approach faces great difficulties in the setting of RAR \ldots  Use of RAR 
eliminates the great majority of standard analysis methods \ldots
\end{displayquote}

\noindent \citet{Rosenberger2016} \secondround{comment on the reason for this:} 

\begin{displayquote}
Inference for [RAR] is very complicated because both the treatment assignments and responses are correlated.
\end{displayquote}


\noindent This raises a key question: how \changes{can} an investigator \secondround{validly} analyze a trial using RAR \changes{in a} frequentist \changes{framework}? \changes{In terms of the notation in Section~\ref{subsec:notation}, this can be formalized as determining whether standard test statistics $T_n$ can be relied on for \secondround{hypothesis testing} (i.e.\ without inflation of type~I error rates), and whether standard estimators $\hat{\theta}_k$ are biased (and if so, by how much).} \changes{Such questions \secondround{are important for adaptive trial designs in general and not only for those using RAR}}. The challenge of statistical inference (within the frequentist framework) is \secondround{naturally} still seen as a \secondround{key} barrier to the use of RAR in clinical practice. \secondround{We next discuss how} valid statistical inference, especially in terms of type~I error rate control and unbiased estimation, is possible for a wide variety of RAR procedures. Note that in what follows, we do not consider time trends and patient drift, \thirdround{as} a separate discussion is given in Section~\ref{subsec:time_trends}. \\

\noindent \changes{\textit{Asymptotic inference}}\\
A straightforward approach to frequentist inference for a trial using RAR is to use standard statistical tests and estimators \textit{without} adjustment. 
This is justified by asymptotic properties that hold for a large class of RAR procedures, including in the multi-arm setting. Firstly, \citet{Melfi2000} proved that an estimator \changes{$\hat{\theta}_k$} that is consistent \changes{(i.e.\ $\hat{\theta}_k \to \theta_k$ as the sample size $n \to \infty$)} when \changes{$Y_{k,i}$} are independent and identically distributed will also be consistent for any RAR procedure \changes{for which $N_k \to \infty$}.

Secondly, \citet{Hu2006} showed that when responses \changes{$Y_{k,i}$} follow an exponential family, simple conditions on the RAR procedure ensure the asymptotic normality of the MLE. \changes{The condition is that} the allocation proportions for each arm \changes{$\sum_{i=1}^n 1\{a_{i,k}=1\}/n \to \rho$, where $\rho \in (0,1)$. This} implies that the RAR procedure \secondround{cannot `select' a treatment during the trial by having allocation probabilities tending to 1 or 0}. Since many test statistics are functions of the MLE, this also implies that the asymptotic null distribution of such test statistics is not affected by the RAR. \changes{Furthermore, if a given RAR procedure does not have this property, then there is a straightforward modification to ensure it holds by bounding (or `clipping') the allocation probabilities~$\pi_{i,k}$, see~\citet{Antognini2021a}}. 
These asymptotic results are the justification for the first guideline given by~\citet{Hu2006} on RAR procedures, which states that ``Standard inferential tests can be used at the conclusion of the trial.'' \\

\noindent \changes{\textit{\thirdround{Finite} sample inference}} \\
\secondround{The validity of} asymptotic results to use standard tests and estimators 
\secondround{requires} a sufficiently large sample size, and the effect of a smaller sample size on inference is greater the more \changes{imbalanced} the RAR procedure is (e.g. see the results in~\citet{Williamson2020})\thirdround{.}
As noted by~\citet{Rosenberger2012}, for some RAR procedures in \secondround{a two-arm setting, there is} extensive literature \secondround{on} the accuracy of \secondround{asymptotic} approximations under moderate sample sizes using simulations~\citep{Hu2003, Rosenberger2004, Zhang2006}. 
For the DBCD, sample sizes of $n = 50$ to $100$ are sufficient, while for urn models reasonable convergence is achieved for $n = 100$. For these procedures, \citet{Gu2010} explored which asymptotic test statistic to use for a clinical trial with a small to medium sample size and binary responses.

When the asymptotic results above cannot be used, either because of small sample sizes or because the conditions on the RAR procedures are not met, then alternative methods for testing and estimation have been proposed. We summarize the main methods below, concentrating on type~I error rate control and unbiased estimation.


A common method for controlling the type~I error rate, particularly for BRAR procedures, is a simulation-based calibration approach, \changes{see e.g. see the FDA guidance on simulations for adaptive design planning~\citep[Section VI.A]{FDA2019}}. Given a trial design that uses RAR and an analysis strategy, a large number of trials are simulated under the null. Applying the analysis strategy to each of these trial realizations gives a Monte Carlo approximation of the relevant 
error rates \changes{(see Section~\ref{subsec:performance})}. If necessary, the analysis strategy can be adjusted to satisfy type~I error constraints. Variations of this approach have been used in \citet{Wason2014, Wathen2017, Antognini2021b}.
\changes{These approaches} can be computationally intensive, \changes{and there are no guarantees beyond the parametric space explored in the simulations}.

A related approach is to use a \textit{re-randomization} test, also known as \textit{randomization-based inference}. In such a test, the outcomes \changes{$\bm{y}^{(n)}$ are taken} as fixed, but the \changes{allocations $\bm{a}^{(n)}$ are} regenerated many times using the RAR procedure \changes{under the null hypothesis}. For each replicate, the test statistic~\changes{$T_n$} is recalculated, and a consistent estimator of the $p$-value is given by the proportion of test statistics that are at least as extreme \thirdround{as} the value actually observed. Intuitively, this is valid because under the null hypothesis of no treatment differences, \changes{$\bm{y}^{(n)}$ and $\bm{a}^{(n)}$} are independent. \citet{Simon2011} give \changes{general} conditions under which the re-randomization test guarantees the type~I error rate \changes{for all RAR procedures}. \citet{Galbete2016a} showed that $15,000$ replicates are sufficient to accurately estimate even very small $p$-values. An advantage of re-randomization tests is that they protect against \textit{unknown} time trends \secondround{(see Section~\ref{subsec:time_trends})}. However, re-randomization tests can suffer from a lower power compared with using standard tests~\citep{Villar2018}, particularly if the RAR procedure has allocation probabilities that are highly variable~\citep{Proschan2019}.

\changes{The implementation} of these methods may \changes{lead to} \secondround{computational cost and Monte Carlo error concerns}. There have been a few proposals that do not rely on simulations. \citet{Robertson2019, Glimm2022} proposed a re-weighting of the usual \thirdround{$Z$}-test that guarantees familywise error control for a large class of RAR procedures for multi-arm trials with normally-distributed outcomes, although with a potential loss of power. \citet{Galbete2016b} derived the exact distribution of a test statistic for a family of RAR procedures in the context of a two-arm trial with binary outcomes, and hence showed how to obtain exact $p$-values.


\secondround{Turning now to estimation bias, the MLEs for the parameters of interest for a trial using RAR will typically be biased in small samples. This is illustrated }
for a number of RAR procedures for binary outcomes through simulation in~\citet{Villar2015a, Thall2015}. However, the latter point out that in their setting, which incorporates early stopping, 

\begin{displayquote}
\ldots most of the bias appears to be due to continuous treatment comparison, rather than AR \textit{per se}.
\end{displayquote}

\noindent Hence it is important to distinguish bias induced by early stopping from that induced by the 
RAR procedure.
\secondround{In a binary setting and for multi-arm RAR procedures without early stopping, the bias of the MLE~\changes{$\hat{p}_k$}} is given in~\citet{Bowden2017}: 
\begin{equation}
\label{eq:bias}
\changes{\text{bias}(\hat{p}_k) = E(\hat{p}_k) - p_k = -\frac{\text{Cov}(N_k, \hat{p}_k)}{E(N_k)}}.
\end{equation}
In \thirdround{a} \secondround{typical} RAR \thirdround{procedure that} assigns more patients to treatments that appear superior \changes{(i.e.\ $\text{Cov}(N_k, \hat{p}_k) >0$)}, \secondround{equation~\eqref{eq:bias}} shows the bias of the MLE is negative. 
The magnitude of this bias is decreasing with the expected number of patients assigned to the treatment \changes{(i.e.\ as $E(N_k)\to \infty$)}. When estimating the treatment \textit{difference} however, the bias can be either negative or positive, which agrees with the results in~\citet{Thall2015}.

\citet{Bowden2017} showed that if there is no early stopping, the magnitude of the bias tends to be small for the RPW rule and the BRAR procedure proposed by~\citet{Trippa2012}. For more \changes{imbalanced} RAR procedures, the bias can be larger however, e.g. see~\citet{Williamson2020}. As a solution, \citet{Bowden2017} proposed using inverse probability weighting and Rao-Blackwellization to produce unbiased MLEs, although these can be computationally intensive. For urn-based RAR procedures, \citet{Coad2001} also proposed bias-corrected estimators. 
For sequential maximum likelihood procedures and the DBCD, \cite{Wang2020} evaluate the bias issue and propose a solution. \secondround{Meanwhile, \citet{Marschner2021} proposed a general framework for analysing adaptive experiments, included trials using RAR, and explored the merits of both conditional and unconditional estimation.}

Finally, adjusted confidence intervals for RAR procedures have received less attention in the literature. \citet{Rosenberger1999} proposed a bootstrap procedure for multi-arm RAR procedures with binary responses, while \citet{Coad2000} proposed corrected confidence intervals for a sequential adaptive design in a two-arm trial with binary responses. Recently, \citet{Hadad2021} proposed a strategy to construct asymptotically valid confidence intervals for a large class of adaptive experiments (including RAR). \\

\noindent \textit{Summary}

For trials with sufficiently large sample sizes, asymptotic results justify the use of standard tests and frequentist inference procedures when using many types of RAR. When asymptotic results do not hold, inference does become more challenging compared with using \changes{ER} \secondround{but it is possible. There is a growing body of literature demonstrating how a trial using RAR, if \thirdround{designed and} analyzed appropriately, can control the type~I error rate and correct for the bias of the MLE}. 
All this should give increased confidence that the results from a trial using RAR can be both valid and convincing.
We reiterate that from a Bayesian viewpoint, the use of RAR does not pose additional inferential challenges. \\

\subsection{\changes{Does using RAR make robust inference difficult} if there is potential for time trends? 
}
\label{subsec:time_trends}

\secondround{The occurrence} of time trends caused by changes in the standard of care or by patient drift (i.e.\ changes in the characteristics of recruited patients over time) is seen as 
a major barrier to the use of RAR in practice

\begin{displayquote}
One of the most prominent arguments against the use of \changes{[RAR]} is that it can lead to biased estimates in the presence of parameter drift. \citep{Thall2015}
\end{displayquote}

\begin{displayquote}
A more fundamental concern with adaptive randomization, which was noted when it was first proposed, is the potential for bias if there are any time trends in the prognostic mix of the patients accruing to the trial. In fact, time trends associated with the outcome due to any cause can lead to problems with straightforward implementations of adaptive randomization. \citep{Korn2011a}
\end{displayquote}

\noindent Both papers cited above show (for BRAR procedures) that time trends can \secondround{substantially} inflate the type~I error rate when using standard analysis methods, and induce bias into the MLE. Further simulation results 
are given in~\citet{Jiang2020}. \citet{Villar2018} present a simulation study for different time trend assumptions and a variety of RAR procedures in trials with binary outcomes including the multi-arm setting.

\changes{As an illustrative numeric example from~\citet{Villar2018}, consider a two-arm trial with binary outcomes, where $n = 100$ and patients are randomized in groups of size~10. Suppose there is a linear upward trend in $p_0$, so that the overall time trend within the trial \[
D = Pr(Y_{0,i} = 1 \, | \, 90 < i \leq 100) - Pr(Y_{0,i} = 1 \, | \, 0 < i \leq 10)
\] varies in $D \in \{0, 0.01, 0.02, 0.04, 0.08, 0.16, 0.24\}$. In this case, \thirdround{under the null scenario where $p_0=p_1$ at all time points}, the optimal allocation of~\citet{Rosenberger2001} has an almost constant type~I error rate, just above the nominal 0.05 level. The TW procedure~\citep{Thall2007} has an inflated type~I error rate (about 0.09) even without any time trend (i.e.\ $D = 0$), which increased to almost 0.15 when $D = 0.24$. Finally, the patient-benefit oriented FLGI rule~\citep{Villar2015b} has a type~I error rate going from 0.05 to almost 0.25 as $D$ increased from 0 to 0.24.} \secondround{These results show that for RAR procedures, even changes in just $p_0$ (or $p_1$) over time can have a considerable impact on operating characteristics. Hence time trends in the treatment effect (however defined) will also be expected to have similar impacts.}

Although time trends can inflate the type~I error when using RAR procedures, there are two important caveats given in~\citet{Villar2018}. Firstly, 
certain power-oriented RAR procedures appear to be effectively immune to \secondround{the} time trends \thirdround{considered in their paper}. In particular, RAR procedures that protect the allocation to the control arm are particularly robust. \changes{A possible explanation is that those rules have a smaller imbalance, as suggested in~\citet{Antognini2021a}.}
\noindent Secondly, \secondround{as discussed in~\citet{Villar2018}}, a largely ignored but highly relevant issue is the size of the trend and its likelihood of occurrence in a specific trial:

\begin{displayquote}
\ldots the magnitude of the temporal trend necessary to seriously inflate the type~I error of the patient benefit-oriented RAR rules need to be of an important magnitude (i.e.\ change larger than 25\% in its outcome probability) to be a source of concern.
\end{displayquote}

A more general issue around time trends is that they can invalidate the key assumption that observations about treatments are exchangeable (i.e.\ that subjects receiving the same treatment arm have the same probability of success). This, in turn, invalidates commonly used frequentist and Bayesian models, and hence the inference of the trial data. Type~I error inflation and estimation bias can be seen as examples of this wider issue.

As~\cite{Proschan2020} put it, temporal trends are  likely to occur in two settings:

\begin{displayquote}
\ldots 
1) trials of long duration, such as platform trials in which treatments may continually be added over many years and 2) trials in infectious diseases such as MERS, Ebola virus, and coronavirus.
\end{displayquote}

\noindent Despite this, little work has looked at estimating these trends, especially to inform trial design in the midst of an epidemic. Investigating these points is essential 
to make a sound assessment of the value of using RAR. \secondround{A recent exception 
is in~\citet{Johnson2021}, where a two-arm vaccine trial for COVID-19 using RAR is studied using a model to simulate the epidemic (including linear trends).} 


As mentioned in Section~\ref{subsec:power}, a \secondround{robust} method to prevent 
type~I error inflation 
is to use a re-randomization test. 
Simulation studies illustrating \secondround{the use of this test can be found} in~\citet{Galbete2016b, Villar2018, Johnson2021}. However, this can come at the cost of a considerably reduced power compared with using an unadjusted testing strategy. \changes{More recently, \citet{Wang2021} showed how to construct confidence intervals for randomization tests that are robust (in terms of coverage) to time trends.}

An alternative to randomization-based inference is to use a stratified analysis. This was first proposed by~\citet{Jennison2000} for group-sequential designs, with subsequent work by~\citet{
Karrison2003, Korn2011a}. These papers show that a stratified analysis can eliminate the type~I error inflation induced through time trends. However, \citet{Korn2011a} also showed that this strategy 
can reduce the trial efficiency \secondround{(see also \citet{Korn2022} for similar arguments)}, \secondround{both} in terms of increasing the required sample size and the chance of patients being assigned to the inferior treatment. 

Another approach is to explicitly incorporate time-trend information into the regression analysis. \changes{\citet{Jennison2001} developed theory that allows the incorporation of polynomial time trends as covariates in a general normal linear regression model for group sequential designs, while} \citet{Coad1991} modified a class of sequential tests to incorporate a linear time trend for normally-distributed outcomes. Meanwhile, \citet{Villar2018} assessed incorporating the time trend into a logistic regression (for binary responses), and showed that this can alleviate type~I error inflation if the trend is correctly specified
However, this \thirdround{leads} to a loss of power and \thirdround{complicates} estimation (due to the technical problem of separation).

Finally, it is possible to try \secondround{to} control the impact of a time-trend \textit{during} randomization. \citet{Rosenberger2001b} proposed a \secondround{CARA} 
procedure for a two-armed trial that can take a specific time trend as a covariate. More recently, \citet{Jiang2020} proposed a \changes{BRAR} procedure that includes a time trend in a logistic regression model, and uses the resulting posterior probabilities as the basis for the randomization probabilities. This model-based procedure controls the type~I error rate and mitigates estimation bias, but at the cost of reduced power.\\[-6pt]

\noindent \textit{Summary}

Large time trends can inflate the type~I error when using RAR, \changes{and this inflation becomes worse the more imbalanced \secondround{the RAR procedure is}}. 
However, 
RAR procedures that protect the allocation to the control arm or \thirdround{impose restrictions to avoid extreme allocation probabilities} are particularly robust. For other RAR procedures, \changes{analysis} methods 
exist to mitigate the type~I error inflation caused by time trends, although \secondround{with} a loss in power. Finally, \secondround{we} note that time trends can affect inference in all types of adaptive clinical trials, and not just those using RAR. \\


\subsection{\changes{Is RAR more challenging to implement in practice?}}
\label{subsec:practical}

\secondround{In addition to the statistical aspects discussed in Sections~\ref{subsec:imbalance}--\ref{subsec:time_trends}}, there \changes{are practical questions} to \changes{consider} to best implement RAR in the context \secondround{of the study} at hand. \changes{Most of these practical issues} \secondround{apply to}  other randomized designs (both adaptive and non-adaptive), so we focus here on a few that \secondround{merit a specific discussion for RAR}. \\

\noindent \textit{Measurement/classification error and missing data}

Measurement error (for continuous variables) or classification error (for binary variables) and missing data are common in medical research. \secondround{There are many approaches proposed to} reduce the impact of these on statistical inference (see e.g.\ \citet{Guolo2008, Little2002, Blackwell2017}) but \secondround{very little} literature on this 
\secondround{in the context of RAR}. The \secondround{distinctive concern} 
is that the sequentially updated allocation probabilities may be biased, \changes{and hence the design will not have its expected properties e.g.\ in terms of patient benefit}. 

\secondround{
A few articles looking at classification (or measurement) error \changes{in RAR} include \citet{Li2012}, who derive optimal allocation targets under 
constant misclassification probabilities that differ between the arms, and \citet{Li2013}, who explore through simulation the effect of misclassification (in the two-arm setting) on optimal allocation designs.}


As for missing data, \thirdround{\citet{Chen2022}} consider the performance of BRAR procedures
under the assumption of missing at random (see \citet{Rubin1976}) and with a single imputation for the missing responses. They found that \thirdround{these procedures encourage more assignments in the arm with missing data, and that simple mean imputation can largely mitigate this effect}.
%
\citet{Williamson2020} 
propose an imputation method for \secondround{a bandit-based RAR} when the outcome is undefined.
Incomplete data for such extreme cases \secondround{is imputed} with random samples drawn from the tails of the distribution. \secondround{Simulations} suggest that imputing in this way is better than \secondround{ignoring missingness} in terms of patient benefit and other metrics.
More complex scenarios, e.g.\ data not missing at random, remain unexplored, but this is the case \changes{for adaptive trials in general} except for some simple settings (see e.g.~\citet{Lee2018}). \\

\noindent \textit{Delayed responses \changes{and recruitment rate}}

The use of RAR is not \changes{feasible} 
\secondround{if} the patient outcomes are only observed after all patients have been recruited and randomized. This \secondround{is rare but} may happen if the recruitment period is \changes{short} (e.g.\ due to a high recruitment rate), or when the outcome of interest takes a long time to observe. One way to address the latter is to use a surrogate outcome that is more quickly observed \secondround{as for example in}
\citet{Tamura1994}. 
Another possibility 
is to use a randomization plan that is implemented in stages as more data becomes available \changes{(like for FLGI)}. %

In general, 
as stated in~\citet[pg. 105]{Hu2006}:

\begin{displayquote}
From a practical perspective, there is no logistical difficulty in incorporating delayed responses into the \secondround{RAR} procedures, provided some responses become available during the recruitment and randomization period. 
\end{displayquote}

\noindent
However, statistical inferences at the end of the trial can be affected. This is explored theoretically for urn models~\citep{Bai2002, Hu2004b, Zhang2007b} as well as the DBCD~\citep{Hu2008}. These papers show that the asymptotic properties of these RAR procedures are preserved under widely applicable conditions. In particular, when more than 60\% of responses are available by the end of the recruitment period, simulations show that the power of the trial is essentially unaffected. \\

\noindent \textit{Patient consent \changes{to be randomized}}

Patient consent protects patients' autonomy, and requires an appropriate balance between information disclosure and understanding~\citep{Beauchamp1997}.
There is evidence that the basic elements to ensure informed consent (recall and understanding) can be difficult to ensure even for non-adaptive studies~\citep{Sugarman1999, Dawson2009}. 
The added complexity of allocation probabilities that may change in response to accumulated data only makes achieving patient consent more challenging.  
Moreover, since these novel adaptive procedures are still rarely used, there is little practical experience to draw upon. 
%
\\

\noindent \textit{Implementing randomization changes \changes{during a study}}

Randomization of patients, whether adaptive or not, must be done in accordance with standards of good clinical practice. As such, in most clinical trials randomization is done through a dedicated and secure web-based system that 
is available 24/7.
%
In the UK, for example, most clinical trials units will outsource their randomization to external companies. 
This 
outsourcing \secondround{is practical but costly, and limits} the ways in which randomization can be implemented to those currently offered by \changes{such} companies. To the best of the authors' knowledge, in the UK \secondround{common providers treat} 
every change in a randomization ratio as a trial change (which is charged as such), rather than being considered an integral part of the trial design. Beyond the extra costs \changes{and limitations to the use of RAR} that this brings, it \secondround{also} introduces unnecessary delays as randomization 
is stopped while the change is implemented.  

A related issue is that of preserving treatment blinding, which is key to the integrity of clinical trials. This is particularly important when \changes{using} RAR, as if an investigator \changes{knows which} treatment is more likely to be allocated next, selection bias is more likely to occur.
In most cases, preserving blindness will require an independent statistician (which requires extra resources) to handle the interim data and implement the randomization, or a data manager can provide data to an external randomization provider who can then update the randomization probabilities independently of the clinical and statistical team. Further discussion on these issues 
can be found in~\citet{Sverdlov2013b}. \\



\subsection{Is using RAR in clinical trials (more) ethical?} 
\label{subsec:ethics}

Ethical reasons \changes{are} the most cited arguments in favor of using RAR \changes{to design clinical trials}.

\begin{displayquote}
Our explicit goal is to treat patients more effectively, but a happy side effect is that we learn efficiently. \citep{Berry2004}
\end{displayquote}

\begin{displayquote}
Research in [RAR] developed as a response to a classical ethical dilemma in clinical trials. \citep{Hu2006}
\end{displayquote}


\noindent Nevertheless, there are also arguments that RAR may not be \changes{ethically preferred}.

\begin{displayquote}
For RCTs \thirdround{[Randomised Controlled Trials]} where treatment comparison is the primary scientific goal, it appears that in most cases designs with fixed randomization probabilities and group sequential decision rules are preferable to AR \secondround{[RAR]} scientifically, ethically and logistically \citep{Thall2016}
\end{displayquote}

Clinical research
poses several ethical \changes{challenges}. There is an inevitable tension between clinical research and
clinical practice, as the latter is concerned with best treating an individual patient \changes{while the former is focused on \secondround{`future'}} patients. 
\changes{Clinical research is associated with a clinical trial whose main aims are the \secondround{testing} and estimation goals as in Section~\ref{subsec:performance}. Clinical practice is directly concerned with patient benefit goals which are, at best, secondary aims in traditional clinical trials}. Such ethical \changes{questions} are becoming more discussed as \changes{personalized} treatment becomes more \changes{embedded into} research, as \changes{is the case for oncology}~\citep{London2018}.

Although 
treating patients \changes{in the trial} ``more effectively" using RAR appears to be ethically attractive, particularly from \changes{the recruited patients' perspective}, the extent to which these \changes{and other} adaptive designs are more ``ethical'' than traditional designs is only starting to be addressed \changes{by ethicists}.
Thus, we do not aim to answer the question whether RAR is (more) ethical or not, as this requires a specific answer for each method and trial context. Instead, we review key concepts that could affect this answer and that come from formal discussions by ethicists. \\

\noindent \emph{The ``equipoise'' concept \changes{and the ethical grounds for randomizing patients} }

Equipoise is \changes{typically defined as} a state of uncertainty of the individual investigator regarding the relative merits of interventions for a population of patients. Such uncertainty justifies randomizing patients to treatments as this does not imply knowingly disadvantaging patients. This \secondround{concept may} 
extend to 
\secondround{include}
``{honest, professional disagreement
among expert clinicians}'' about the relative merits
of interventions
~\citep{Freedman1987}. 
This broader definition is known as ‘clinical equipoise’ while the \secondround{former} is ‘theoretical equipoise’.

\changes{An argument against the use of RAR is that it violates the principle of equipoise on which clinical trials is based upon~\citep{Laage2017}.}
Changing the randomization probabilities
in light of patients' responses \changes{may be} viewed as \changes{breaking} equipoise, because the updated allocation weights reflect the relative performance of the interventions in question. Once the randomization weights become unbalanced,
the study has a preferred treatment and allocating participants
to treatments regarded as inferior \changes{could be considered} unethical. 
However, this argument that RAR is \emph{unethical} because it breaks equipoise
is based on two assumptions:  1) randomization ratios reflect a single agent's beliefs about the relative merits of the interventions being tested; 
and 2) equipoise is a state of belief in which the relevant probabilities are assumed to be equally balanced. Neither of these two assumptions are consistent with the definition of `clinical equipoise' as the clinical community is \secondround{multi-}agent and disagreement among these agents will not necessarily correspond to a 50\%-50\% split of opinions.  \\

\noindent
\emph{Patient horizon (individual and collective ethics)}

The ethical value of RAR (and of other \changes{trial} designs) depends \changes{directly on the trial's specific aim in relation to its context. For example, a feature that considerably affects comparisons of design options is disease prevalence} \secondround{(a concept} 
linked to that of patient horizon~\citep{Anscombe1963, Colton1963}). Suppose a clinical trial is being planned where~$T$ denotes
the ``patient horizon'' for that study, i.e.\ those patients within and outside of the trial who will benefit from its conclusions.
The \secondround{exact} value of~$T$ is never known but \changes{its order of magnitude considerably impacts the relative merits of competing trial goals}. A trial \changes{relevant to patients with coronary artery disease will have the vast majority of the patient horizon outside of the trial, making the inferential goals of the study of paramount importance. On the other hand, a rare pediatric cancer is likely to have a large proportion of the patient population in the trial, heightening the tension between patient benefit and inferential goals}. Similar considerations apply for emerging life-threatening diseases (e.g.\ the Ebola outbreak or the COVID-19 pandemic), where the patient horizon can be short for reasons other than prevalence. 
\secondround{When} the choice of design is based only on inferential considerations, there will be many instances in which a design may be considered inferior from a patient benefit viewpoint.

The impact of \changes{$T$} on the \changes{ethical} comparison of designs depends on considerations around individual and collective ethics and potential conflicts between these two. As \citet[pg. 775]{Tamura1994} express it, RAR ``represents a middle ground between the community benefit and the individual patient benefit'' and because of this ``it is subject to attack from either side''. 
This point has been well discussed and formally studied in the statistical literature (see \citet{Berry1995, Cheng2003, Berry2004}). Despite this, prevalence of a disease is almost never taken into account, neither in practice when designing trials nor in \changes{methodological} articles comparing RAR from an ethical point of view. \secondround{See} \citet{Lee2021, Metelkina2017} for \changes{recent attempts to address this}. \\



\noindent
\emph{Summary}

\secondround{We believe that} the ethics of RAR \secondround{needs} more attention from ethicists, \secondround{including} collaborations between ethicists and statisticians to address the caveats and complexities of this broad family of methods. Positions based purely on statistical or ethical arguments in isolation are likely to be \secondround{inadequate and arguments that involve ethical metrics should ideally be jointly discussed with multiple stakeholders. 
It is important to bear in mind that}
compromises between statistical and ethical objectives have very different implications under different settings. \thirdround{For example, the trade-offs between these two objectives may look very different in a two-arm trial setting compared to a multi-arm trial.}

\changes{Ideally, how this interaction between ethics and statistics can proceed is as follows (as suggested by an anonymous reviewer). \secondround{Ethics informs the relative importance of a trial's goals, in particular the} balance between individual and collective benefit. \secondround{Once these priorities are in place}, a statistical design that achieves these goals
can be proposed.
The ethical aspects can be revisited in light of the resulting properties of the statistical design. For example, suppose RAR is chosen to deliver a certain level of benefit to patients in the trial. This may require an increase in the trial size to preserve the inferential properties for future patients to \secondround{be} ``ethical''. In that case, depending on the prevalence of the disease and the general context, a larger trial using the original RAR procedure may still deliver the most benefit to all patients and remain the preferred option. If this is not the case, then the ethics-design choice \secondround{can be} revisited.}\\

\section{\protect\secondround{Final considerations and discussion}}
\label{sec:discuss}

\secondround{The pace of methodological work on RAR and the debate over its use has certainly sped up in recent years, driven by the response to challenges during health crises like the COVID-19 pandemic and the increase uptake of these methods in machine learning and data science more generally. However, to some extent, the debate and methodological progress remain disconnected from each other.   
\changes{It is important to} bear in mind that generalizations within such a large class of methods \changes{run the risk of being} partial and misleading. \thirdround{Even for a single RAR procedure, its performance may vary considerably across the parameter space of interest}. \secondround{In this paper we have aimed} to illustrate the breadth of RAR procedures \secondround{by presenting a critical (but balanced) appraisal of well established views about RAR, and to help guide future research efforts towards areas that have received less attention}.} 




\changes{We emphasize that this paper does not advocate 
for the use of RAR in all trial settings \thirdround{(but we also do not intend to discourage trialists from considering its use in general). }
There are contexts where other trial adaptations or even \thirdround{a fixed randomization} design may be preferable for both methodological or practical reasons. This is important to consider with adaptive trials in general -- sometimes it may be better to `keep it simple' and use traditional non-adaptive designs instead~\citep{Wason2019}.}
\changes{However, when \thirdround{the} use of RAR is considered, it is helpful to remember that RAR encompasses a large set of possible design \thirdround{(and analysis)} options, rather than being a homogeneous technique to either include or not.} Indeed, many of the recent general criticisms and praise for \changes{RAR} in clinical trials has been driven by arguments that apply to \secondround{the} particular subclass \secondround{of BRAR}, but may well not be as relevant for other RAR procedures.
%

\changes{Trade-offs \secondround{in terms of different metrics} are ubiquitous and in many cases unavoidable \secondround{in clinical trials},
\secondround{as} RAR procedures can address a specific need at the expense of a cost in a different area.
Hence, a RAR procedure should be chosen carefully according to the specific context and goals of a trial, in light of the practical challenges and constraints that implementing RAR poses.} 
%
%
\changes{
Indeed, as noted by an anonymous reviewer, the approach of starting with a set of different RAR procedures \secondround{and then choosing one based on comparing their performance as measured by different metrics}
is \secondround{arguably} going in the wrong direction. Instead, a preferable approach is to explicitly start by defining the type of trial and the investigators’ priorities in setting goals for the trial, and to then select a RAR design suited to these goals \thirdround{(see also~\citet{Pitt2021})}}.

\secondround{Starting with the type of trial, factors such as the phase of clinical development, the number of treatment arms 
and the clinical endpoint will naturally influence the aims of the trial and the appropriateness of a design including RAR. Some types of clinical trial may be particularly suited to the use of a well-chosen RAR procedure -- for example,} in multi-arm trials it is natural to consider dropping poorly performing treatment arms, and RAR offers an intermediate option of reducing numbers \secondround{on such} treatments.

\secondround{Given a particular type of trial, the aims of the trial can then be considered. Broadly speaking, these aims} fall into 
two categories:
\begin{enumerate}[label = \arabic*)]
    \item \changes{Determine how best to treat future patients after the trial concludes while avoiding (or minimizing) harm to patients in the trial;}
    \item \changes{Optimize treatment of patients in the trial itself (i.e.\ treat patients in the trial as effectively as possible).}
\end{enumerate}
Depending on the relative importance of these two,
different RAR rules may be appropriate.

\secondround{Once the aims of a study, their relative importance, and the corresponding metrics have all been agreed upon, the question of what an optimal RAR procedure is in terms of those metrics can be addressed. Ideally (as suggested by an anonymous editor), an optimal trial design can be found within this framework, rather than proposing ad hoc procedures and testing them against different metrics. However, in the literature reviewed we found} \changes{the use of the term `optimal' in relation to RAR procedures \secondround{can have} many different meanings. 
A broader definition of \emph{optimality} \secondround{may be beneficial to consider}, \secondround{not only including} optimal allocation targets but \secondround{also} RAR families that have some other form of optimality (or near optimality). 
\secondround{In any case, it is important} to explicit say in what sense a procedure is `optimal' when using this terminology.}


\changes{As a general point (and one we more fully appreciate following recent discussions with applied trial statisticians), it is crucial to not consider statistical or methodological issues in isolation of practical issues. This may be key for the design of any experiment, but is more important for RAR and adaptive designs in general. 
For example, selection bias may be a big issue in some contexts, and if blinding is not possible, then the use of RAR may be less appropriate. Hence, greater collaboration and discussion between methodologists and applied trialists \secondround{is useful} to ensure that methods are developed with practical considerations in mind.}
    

\secondround{We would like to end with a short summary as to what we feel the future for RAR methods research should bring to improve its usefulness in clinical practice. We wrote this paper in a attempt to reconcile conflicting perspectives as much as to} %
motivate researchers to address \secondround{the} issues \changes{mentioned here} with new ideas. \secondround{New work is needed} to realise most of the \changes{potential} advantages of RAR 
with fewer of its downsides while taking the trial context into account. \changes{With the increasing use of response-adaptive procedures in machine learning and data science more generally, this presents a golden opportunity for biostatisticians to embrace and lead the development of this wide adaptive class in both theory and practice.}

\changes{As a general point, \secondround{our hope is that any contribution to} RAR methodology \secondround{should be well contextualised} within the ongoing debate \secondround{in order to achieve practical impact and to avoid repeating common arguments that are already well-represented in the literature. 
} When developing new proposals, it can be helpful to define terminology carefully, report a wide range of metrics and to be explicit about the potential limits of the conclusions made.}


\secondround{Firstly, we encourage the explicit definition and clear reporting of} the metrics used to evaluate RAR procedures, 
%
\secondround{as well as a} broad look at multiple metrics (not just standard 
operating characteristics).
\secondround{For example}, estimation \secondround{and sample size imbalance} metrics are relatively under-reported in the literature. 
Similarly, since many RAR procedures \secondround{impact} patient benefit, including at least one \secondround{such} metric (see 
Section~\ref{subsec:performance}) \secondround{is useful} when comparing RAR procedures. 
%

\secondround{Exploring a wide parametric space in simulations (and not only subsets of interest) can also be key}. \secondround{For example}, Neyman allocation maximizes power, but for $p_0 + p_1 > 1$ assigns more patients to the inferior arm (see Section~\ref{subsec:power}). Similarly, 
for the RPW rule the limiting distribution of the allocation proportion depends on whether $p_0 + p_1 > 3/2$~\citep{Rosenberger2016}.
\secondround{Given the above, it is also important to discuss when certain properties} may not apply to other RAR families. This could reduce the chances of readers misunderstanding the scope of conclusions about a specific family of RAR procedures. More generally, definitive statements based only on simulation results should be regarded with an appropriate degree of caution.  There are no universal set of rules on how to conduct simulation studies (although useful guidelines are proposed by~\citet{Morris2019}). 




\changes{
\secondround{In terms of specific methodological research areas, a key open area} is that of efficient and valid inference methods for RAR. As discussed in Section~\ref{subsec:stat_inference}, \secondround{a simple asymptotic approach for inference is valid in many case but it does} not apply to all RAR \thirdround{procedures}. 
On the other hand, 
valid methods 
for small samples (or time trends) \secondround{such as randomization-based inference} suffer from low power. Hence, \secondround{the development of} new inferential procedures for \thirdround{finite} samples \secondround{that} do not suffer from a large loss in power \secondround{would be very useful} (as a recent example along these lines, see~\citet{Barnett2020, Deliu2021}). \secondround{For time trends in particular,} there has been little work estimating the likelihood and magnitude of such trends in practice, especially in contexts such as emerging epidemics. More research would help to determine whether RAR would be appropriate for specific trial contexts.}

\secondround{Another open research question is} how to account for missing data or measurement error when using RAR. Adjusted confidence intervals have also received \secondround{little} attention in the literature. \secondround{More generally, further work is needed to expand the comparison of multi-arm RAR procedures (particularly in terms of different power definitions) beyond BRAR.}
In terms of design aspects, RAR has the under-explored potential for addressing delicate issues when designing studies with composite or complex endpoints. \secondround{Another consideration is that block-randomized versions of RAR methods are much more likely to be applied in practice than fully sequential schemes, but open questions remain about how these implementations compare in terms of power and patient benefit}. \thirdround{As well}, it is still unclear in general how trial designs incorporating RAR compare with well-chosen group sequential and MAMS designs.

\thirdround{Finally, regardless of methodological considerations and future development, the use of RAR in practice would stil require the availability of user-friendly software for both the implementation of the randomization algorithm as well as for the analysis approaches that were mentioned in Section~\ref{sec:myths} of this paper.}\\





\section*{Acknowledgments}

\changes{The authors thank the Editor, Associate Editor and the four anonymous reviewers for their constructive comments which helped substantially improve this paper}. We \changes{also} thank Peter Jacko for many helpful comments on an earlier version of this work, Andi Zhang for providing code for the FLGI procedures used in Section~\ref{subsec:imbalance}, Ayon Mukherjee for suggesting the use of the drop-the-loser rule in Section~\ref{subsec:imbalance}, Arina Kazimianec for her work on sample size imbalance which helped motivate Section~\ref{subsec:imbalance}, and Nikolaos Skourlis for screening literature on model-based adaptive randomization. The authors acknowledge funding and support from the UK Medical Research Council (grants \changes{MC\_UU\_00002/15 (SSV)}, MC\_UU\_00002/3 (BCL-K), MC\_UU\_00002/14 (DSR), MR/N028171/1 (KML)), the Biometrika Trust (DSR) and the \changes{NIHR Cambridge Biomedical Research Centre (BRC-1215-20014)}  (DSR, KML, BCL-K, SSV). The views expressed in this publication are those of the authors and not necessarily those of the NHS, the National Institute for Health Research or the Department of Health and Social Care (DHSC). For the purpose of open access, the author has applied a Creative Commons Attribution (CC BY) licence to any Author Accepted Manuscript version arising. \\

\noindent \textbf{Data availability}: \thirdround{Code to implement RAR algorithms given in Section~\ref{subsec:imbalance} can be found at the end of the `Papers and code' section at \url{https://www.mrc-bsu.cam.ac.uk/software/miscellaneous-software/}} \\



\newpage

\section*{Appendix}

\setcounter{table}{0}
\renewcommand{\thetable}{A\arabic{table}}

\begin{table}[ht!]
\caption{Table of acronyms used in the paper. \label{tab:acronyms}}
\vspace{6pt}
\centering
\setlength{\tabcolsep}{12pt}
\renewcommand{\arraystretch}{1.5}

\begin{tabular}{l l l}
\hline
\textbf{Acronym} && \textbf{Definition} \\ \hline
BRAR && Bayesian Response-Adaptive Randomization \\
CARA && Covariate-Adjusted Response-Adaptive \\
DBCD && Doubly-adaptive Biased Coin Design \\
DTL && Drop-The-Loser \\
ENF && Expected Number of Failures \\
ENS && Expected Number of Successes \\
ER && Equal Randomization \\
ERADE && Efficient Response-Adaptive Randomization Designs \\
FLGI && Forward-Looking Gittins Index \\
MAMS && Multi-Arm Multi-Stage \\
MLE && Maximum Likelihood Estimator \\
PBR && Permuted Block Randomization \\
RAR && Response-Adaptive Randomization \\
RCT && Randomized Controlled Trial \\
RPW && Randomized Play-the-Winner \\
TS && Thompson Sampling \\
TW && Thall and Wathen \\

\hline

\end{tabular}

\end{table}

\begin{table}[ht!]
\caption{Properties of various patient allocation procedures, where $p_0 = 0.25$ and $p_1 = 0.45$. Results are from $10^4$ trial replicates. \label{tab:imbalance2} \vspace{6pt}}

\centering
\setlength{\tabcolsep}{12pt}
\renewcommand{\arraystretch}{1.5}

\begin{tabular}{l l l l l}
 \hline
 & & \multicolumn{3}{c}{\textbf{Patient Benefit Metrics}}\\
 $\bm{n}$ & \textbf{Procedure} & $\bm{N_1 - N_0}$  & $\bm{\hat{S}_{0.1}}$ & \textbf{ENS} \\ \hline
 200 & ER & 0 (-28, 28) & 0.069 & 70 (6.8) \\
 & PBR & 0 & 0 &  70 (6.6) \\
 & Oracle & 200 & 0 &  90 (7.0) \\
 & TS & 147 (-54, 194) & 0.032 & 85 (9.5) \\
 & FLGI($b=5$) & 165 (48, 194) & 0.014 &  86 (8.6) \\
 & FLGI($b=10$) & 164 (56, 192) & 0.014 &  86 (8.6) \\
 & TW(1/2) & 124 (24, 182) & 0.008 & 82 (8.2) \\
 & TW($i/2n$) & 88 (20, 148) & 0.001 &  79 (7.6) \\
 & RPW & 30 (-2, 64) & 0.000 & 73 (7.1) \\
 & DBCD & 30 (6, 58) & 0.000 & 73 (6.6) \\
 & ERADE & 29 (8, 52) & 0.000 & 73 (6.6) \\ 
 & DTL & 30 (10, 50) & 0.000 & 73 (7.1) \\ \hline
\end{tabular}

\end{table}

\end{document}